\documentclass[prd,aps,twocolumn,a4paper,floatfix,nofootinbib]{revtex4-2}
\usepackage[utf8]{inputenc}
\usepackage{graphicx,psfrag}
\usepackage{mathrsfs}
\usepackage{amsmath,amsfonts,amssymb}
\usepackage{multirow}
\usepackage{diagbox}
\usepackage{comment}
\usepackage{xcolor}
\usepackage{enumerate}
\usepackage{booktabs}
\usepackage[normalem]{ulem}
\usepackage{hyperref}
\usepackage{mathtools}
\usepackage{siunitx}
\usepackage{float}
\usepackage{textgreek}
\hypersetup{
    colorlinks = true,
    linkcolor = {blue},
    citecolor = {blue},
    urlcolor = {blue},
    linkbordercolor = {white},
    }
\usepackage{color}
\usepackage{bigints}
\definecolor{cyan}{rgb}{0,0.9,0.9}
\definecolor{orange}{rgb}{0.9,0.5,0}
\definecolor{magenta}{rgb}{1,0,1}
\definecolor{purple}{rgb}{0.8,0.4,0.8}
\definecolor{gray}{rgb}{0.8242,0.8242,0.8242}
\definecolor{green}{rgb}{0.,0.8,0.}
\usepackage{subfigure}
\usepackage{tabularx}

\DeclareMathAlphabet\mathbfcal{OMS}{cmsy}{b}{n}

\usepackage[normalem]{ulem}

\begin{document}

\title{On the Role of Muons in Binary Neutron Star Mergers: First Simulations}

\author{Henrique \surname{Gieg}$^{1}$}
\email[E-mail:~]{henrique.gieg@uni-potsdam.de}
\author{Federico \surname{Schianchi}$^{1}$}
\author{Maximiliano \surname{Ujevic}$^{2}$}
\author{Tim \surname{Dietrich}$^{1,3}$}

\affiliation{$^{1}$Institut f\"ur Physik und Astronomie, Universit\"at Potsdam, Haus 28, Karl-Liebknecht-Str. 24/25, 14476, Potsdam, Germany}
\affiliation{$^{2}$Centro de Ci\^encias Naturais e Humanas, Universidade Federal do ABC, 09210-170, Santo Andr\'e, S\~ao Paulo, Brazil}
\affiliation{${^3}$Max Planck Institute for Gravitational Physics (Albert Einstein Institute), Am M\"uhlenberg 1, Potsdam 14476, Germany}

\begin{abstract}
In this work we present a set of binary neutron star (BNS) merger simulations including the net muon fraction as an additional degree-of-freedom in the equation of state (EoS) and hydrodynamics evolution using the numerical-relativity code BAM. Neutrino cooling is modeled via a neutrinos leakage scheme, including in-medium corrections to the opacities and emission rates of semi-leptonic charged-current reactions, although within the elastic approximation. We show that, for our particular choice of baseline baryonic EoS, the presence of muons delays the gravitational collapse of the remnant compared to the case where muons are neglected. Furthermore, when muons and muonic weak reactions are considered, no gravitational collapse occurs within our simulation time and muons are confined in the densest portions of the remnant, while the disk is effectively colder, less protonized and de-muonized. Accordingly, ejecta properties are affected, e.g., ejecta masses are systematically smaller for the muonic setups and exhibit a larger fraction of neutron-rich, small velocity material. Overall, our results suggest that the inclusion of muons and muon-flavored neutrino reactions in the context of BNS merger simulations should not be neglected, thus representing an important step towards more realistic modeling of such systems.
\end{abstract}
\date{\today}
\maketitle

\section{Introduction}
\label{sec:Int}
Merging binary neutron stars (BNS) are among the most prominent sources of multimessenger signals, which combine gravitational-wave (GW) measurements~\cite{TheLIGOScientific:2017qsa, LIGOScientific:2014pky, VIRGO:2014yos} with their associated electromagnetic (EM) counterparts, such as gamma-ray bursts~\cite{Hajela:2019mjy,Hajela:2021faz,Balasubramanian:2022sie, Nedora:2022kjv, Wang:2024wbt, Savchenko:2017ffs, Abbott:GRB} and kilonovae, e.g. AT2017gfo~\cite{LIGOScientific:2017ync, Arcavi:2017xiz, Coulter:2017wya, Lipunov:2017dwd, Tanvir:2017pws, Valenti:2017ngx}.
Such observations allow, for instance, to constrain the uncertain Equation of State (EoS) governing neutron-star matter at supranuclear densities~\cite{Annala:2017llu,Bauswein:2017vtn,Fattoyev:2017jql,Ruiz:2017due,Shibata:2017xdx,Radice:2017lry,Most:2018hfd,Tews:2018iwm,Coughlin:2018miv,Coughlin:2018fis,Capano:2019eae,Dietrich:2020efo,Nedora:2021eoj,Huth:2021bsp, Koehn:2024ape, Koehn:2024set}, assess cosmological properties of the Universe~\cite{Guidorzi:2017ogy,Hotokezaka:2018dfi,Coughlin:2019vtv,Dietrich:2020efo,Perez-Garcia:2022gcg,Wang:2020vgr,Bulla:2022ppy} and to study the formation of heavy elements~\cite{Lattimer:1974slx,Rosswog:1998hy,Rosswog:2005su}.

The interpretation of multimessenger signatures produced during and after a BNS merger requires one to correlate theoretical predictions with the observed signatures. Due to the inherently nonlinear dynamics of such compact systems, numerical-relativity (NR) simulations are required to compute the emitted GWs and predict thermodynamical properties of matter along the post-merger.
In the context of BNS merger simulations, state-of-art comprises attempts to capture, as realistically as possible, a myriad of phenomena that is expected to take place under the extreme conditions encountered throughout the inspiral, merger and post-merger stages. Some examples include accurate modeling of the matter and hydrodynamics~\cite{Shibata:2003ga, Shibata:2005ss, Rezzolla:2010fd, Bauswein:2013yna, Hotokezaka:2013iia, Dietrich:2015iva, Dietrich:2015pxa, Bernuzzi:2016pie, Radice:2013hxh, Radice:2015nva, Radice:2017zta}, the role of magnetic fields ~\cite{Giacomazzo:2014qba,Siegel:2014aaa,Kiuchi:2015sga,Ciolfi:2020cpf,Palenzuela:2021gdo, Kiuchi:2023obe, Neuweiler:2024jae} and the inclusion of neutrino transport~\cite{Ruffert:1995fs, Rosswog:2003rv, Sekiguchi:2010fh, Shibata:2011kx, Foucart:2015gaa, Lehner:2016lxy, Bovard:2017mvn, Foucart:2015vpa,Foucart:2016rxm, Radice:2016dwd, Radice:2018pdn, Radice:2021jtw, Schianchi:2023uky, Radice:2023zlw}. 

One key assumption regarding the modeling of matter is that the EoS contains only electrons $e^-$ and positrons $e^+$ as representative leptonic species. Hence, matter properties are described by EoSs that are represented as three-dimensional functions of the baryon rest-mass density $\rho$, temperature $T$ and net electron fraction $Y_e$.
However, Core-Collapse Supernovae simulations~\cite{Bollig:2017lki, Guo:2020tgx, Fischer:2020vie} show that muons $\mu^-$ and antimuons $\mu^+$ are produced in non-negligible amounts via weak reactions during the formation of a neutron star (NS). Over larger timescales, when the matter cools and reaches the neutrinoless $\beta$-equilibrium, muons are expected to be present wherever the electron chemical potential $\mu_e$ exceeds the muon rest-mass $m_\mu$, i.e., $\mu_e \geq m_\mu c^2\approx 106~\rm{MeV}$~\cite{Shapiro:1983du, Glendenning:1997wn, Haensel:2000, Haensel:2001}. Hence, the description of matter encompassed by the EoS should include an additional degree-of-freedom, accounting for the net muon fraction $Y_\mu$.

Besides, it has been shown that the presence of muons in the interior of neutron stars leads to important microphysical consequences, e.g. the arise of bulk viscosity due to leptonic reactions~\cite{Alford:2010jf, Alford:2021lpp, Harris:2024ssp}, the modification of the direct Urca threshold~\cite{Lattimer:1991ib}, the occurrence of muon-flavored weak reactions and an overall increased proton fraction in locally neutral, $\beta$-stable matter when compared to EoSs that include only $e^-$ and $e^+$.

Interestingly, up to our knowledge, there are only two studies that address the possible impacts of considering muons in BNS systems. The first one is Ref.~\cite{Loffredo:2022prq}, which employs post-processing of data from BNS merger simulations that were produced neglecting the presence of muons. Such a procedure provided important insights about the role of muons during the post-merger stage with respect to the hydrodynamics of matter and the behavior of neutrinos in the trapped regime. However, the method is not able to capture in details the complete dynamics and post-merger evolution of a BNS that is simulated ab-initio including muons and treating the neutrinos-driven interactions on-the-fly.
More recently, Ref.~\cite{Pajkos:2025oyf} performed BNS merger simulations without neutrinos transport, but including pions, muons and trapped neutrinos in the EoS under the assumptions of (i) strong interaction equilibrium among negatively charged pions, neutrons and protons, (ii) weak interaction equilibrium among neutrinos, leptons and nucleons and (iii) a prescription for the muon-flavored number fraction as a function of the rest-mass density. Hence, their modified EoS is 3-dimensional and does not account for the advection of muons. Nevertheless, important dynamical effects are noted, e.g., changes in the thermal and compositional profile of the remnant, in the ejecta properties and in the post-merger GW frequencies.

In this work we intend to surpass the shortcomings of previous investigations and present NR simulations of BNS mergers carried out with the inclusion of muons in the EoS and hydrodynamics, and the use of a Neutrinos Leakage Scheme (NLS) to model the cooling of matter in response to the production of neutrinos, in particular accounting for muon-flavored neutrinos.
The structure of this paper is as follows: in Sec.~\ref{sec:EoS} we describe our procedures for the construction of EoSs including muons and the modifications of the general-relativistic hydrodynamics (GRHD) equations. In Sec.~\ref{sec:NLS} we present details about the NLS implementation and the computation of emissivities and opacities for set of weak reactions. In Sec.~\ref{sec:Methods} we state our methods and BNS setups considered in this worked, which were evolved with the BAM code~\cite{Bruegmann:2006ulg, Thierfelder:2011yi, Dietrich:2015iva,Bernuzzi:2016pie, Gieg:2022, Schianchi:2023uky, Neuweiler:2024jae}. In Sec.~\ref{sec:M-PM-Dyn} we present a qualitative discussion about the merger and post-merger dynamics. In Sec.~\ref{sec:Ej} we make an analysis of the ejecta properties. Finally, in Sec.~\ref{sec:Conc} we state our concluding remarks.
Throughout this work we adopt units in which the gravitational constant $G$, the speed of light in vacuum $c$, the solar mass $M_\odot$ and the Boltzmann constant $k_B$ are equal to one. Greek letters represent spacetime indices ranging from $0$ to $3$, while Latin letters are used for spacelike tensor and range from $1$ to $3$. The spacetime metric $g_{\mu \nu}$ has signature $(-,+,+,+)$.

\section{Equation of State and Hydrodynamics}
\label{sec:EoS}

Generally, a muonic EoS may be constructed by ``dressing'' a baseline baryonic EoS\footnote{Such as those found, for example, in the CompOSE database~\url{https://compose.obspm.fr/}.}, parameterized by the baryon number density $n_b$, temperature $T$, and charge fraction of strongly interacting particles $Y_q$ with a leptonic EoS. In the following, we consider the leptons and the anti-leptons $l = \{e^-, \mu^-, e^+, \mu^+\}$ as relativistic ideal Fermi gases. Hence, the lepton number density $n_{l^{\mp}}$, energy density $\varepsilon_{l^{\mp}}$ and pressure $p_{l^\mp}$ read~\cite{1977ApJ...212..859B, Timmes_1999}
\begin{eqnarray}
    &&n_{l^{\mp}} = K_l \beta_l^{3/2}\left[F_{1/2}(\eta^0_{l^{\mp}}, \beta_l) + \beta_l F_{3/2}(\eta^0_{l^{\mp}}, \beta_l) \right],\label{eq:n-dens} \\
    &&\varepsilon_{l^\mp} = K_l m_l c^2 \beta_l^{5/2}\left[F_{3/2}(\eta^0_{l^{\mp}}, \beta_l) + \beta_l F_{5/2}(\eta^0_{l^{\mp}}, \beta_l) \right] \nonumber \\
    &&\hspace{1cm}+m_l c^2 n_{l^\mp}, \label{eq:e-dens}\\
    &&p_{l^\mp} = \frac{ K_l m_l c^2}{3} \beta_l^{5/2}\left[2 F_{3/2}(\eta^0_{l^{\mp}}, \beta_l) + \beta_l F_{5/2}(\eta^0_{l^{\mp}}, \beta_l) \right], \nonumber \\ \label{eq:press}
\end{eqnarray}
where $m_l$ is the lepton rest-mass, $\beta_l = T/m_l c^2$ is the relativity parameter, $K_l$ is a constant
\begin{equation}
    K_l = 8\pi \sqrt{2} (m_l c^2/hc)^3,
\end{equation}
$\eta^0_{l^\mp}$ are the non-relativistic degeneracy parameters
\begin{eqnarray}
    \eta^0_{l^-} &=& \frac{\mu_{l^-} - m_lc^2}{T},\label{eq:chem-pot-p} \\
    \eta^0_{l^+} &=& -\left(\eta^0_{l^-} + \frac{2}{\beta_l}\right) \label{eq:chem-pot-a},
\end{eqnarray}
and $\mu_{l^-}$ is the relativistic lepton chemical potential (including rest-mass). Note that Eq.~\eqref{eq:chem-pot-p} is a definition, while Eq.~\eqref{eq:chem-pot-a} arises from the equilibrium between particles, antiparticles and photons (with zero chemical potential). Finally, $F_{k}(\eta^0_{l^{\mp}}, \beta_l)$ is the generalized Fermi integral of order $k$, whose evaluation is performed numerically following~\cite{Aparicio_1998}.

The procedure to construct our complete EoS is presented in the following. The first step of our method consists in considering a baseline baryonic EoS parameterized by $(n_b, T, Y_q)$, which is also taken as the baseline 3-dimensional space for the addition of the leptonic sectors of the EoS. It is important to mention that in this work we consider baseline baryonic models containing only protons, neutrons and nuclei. In this case, $Y_q = Y_p$ and we remark that, throughout this manuscript, $Y_p$ stands for the free EoS parameter. Next, from Eqs.~\eqref{eq:n-dens},~\eqref{eq:chem-pot-p} and~\eqref{eq:chem-pot-a} we define the net lepton fraction $Y_l$ as
\begin{eqnarray}
    Y_l (n_b, T, \eta^0_{l^-}) = [n_{l^-} (\eta^0_{l^-}, T) - n_{l^+} (\eta^0_{l^+}, T)]/n_b,\label{eq:yl-def}
\end{eqnarray}
which depends on $(n_b, T)$ and on the non-relativistic lepton degeneracy parameter $\eta^0_{l^{\mp}}$.

Hence, for each pair $(n_b, T)$, $Y_l$ is a function of $\eta^0_{l^-}$ alone. By choosing values of $Y_l$ for a pair $(n_b, T)$, which corresponds to fixing the triple $(n_b, T, Y_l)$, $\eta^0_{l^-}$ may be recovered by numerically inverting Eq.~\eqref{eq:yl-def}, giving $\eta^0_{l^-} = \eta^0_{l^-}(n_b, T, Y_l)$. Consequently, the thermodynamical properties $\xi_l$ associated to the lepton/anti-lepton gas, such as pressure, energy density and entropy, also become functions of $(n_b, T, Y_l)$, as 
\begin{eqnarray}
    \xi_l &=& \xi_{l^-}(T, \eta^0_{l^-}(n_b, T, Y_l)) + \xi_{l^+}(T, \eta^0_{l^+}(n_b, T, Y_l)) \nonumber \\
    &=& \xi_{l^-}(n_b, T, Y_l) + \xi_{l^+}(n_b, T, Y_l).
\end{eqnarray}
Then, for a chosen set of $Y_l$, a tabulated EoS may be produced in the parameter space $(n_b, T, Y_l)$, representing a lepton/anti-lepton species.

More concretely, what we do is to choose a range of muon fractions $Y_\mu = [1\times 10^{-4}, 1\times 10^{-1}]$, logarithmically spaced with $20$ points per decade, plus a $Y_\mu = 0$ slice, for a total of $N_{Y_\mu} = 62$ points. Then we tabulate an EoS for (anti)muons $\xi_\mu(n_b, T, Y_\mu)$ in the same parameter subspace $(n_b, T)$ of the baseline baryonic EoS. Note that the (anti)muon EoS is completely independent of the baryonic sector. Still, it is possible to combine both sectors into one EoS by simply appending one dimension to any of the EoSs parameter spaces. Our choice is to add the $Y_\mu$ degree-of-freedom to the baseline baryonic EoS parameter space such that to each point $(n_b, T, Y_p)$ corresponds $N_{Y_\mu}$ entries, promoting a 3-dimensional EoS to the 4-dimensions $(n_b, T, Y_p, Y_\mu)$.

The next step is to produce a tabulated EoS for the electron/positron gas. Again, following the procedure previously outlined, one could freely choose a range of electron fractions $Y_e$ consistent with $(n_b, T)$ and produce an EoS for electrons/positrons, in principle completely independent of the baryonic and muonic sectors. However, here we choose the $Y_e$ values of our tabulation enforcing charge neutrality at each point of the 4-dimensional EoS parameter space, i.e., for each point $(n_b, T, Y_p, Y_\mu)$ we make
\begin{equation}\label{eq:ch-neut}
    Y_e = Y_p - Y_\mu.
\end{equation}
Then we solve Eq.~\eqref{eq:yl-def} to obtain $\eta^0_{e^-} = \eta^0_{e^-}(n_b, T, Y_p, Y_\mu)$, which then sets the thermodynamical properties of the resulting electron/positron gas $\xi_e(T, \eta^0_{e^-}) = \xi_e(n_b, T, Y_e = Y_p-Y_\mu)$. Finally, we compute the pressure $p$ and energy density $\varepsilon$ of a fluid element by adding all species contributions (as well as contributions from photons $\gamma$), giving
\begin{align}
    p = p_b(n_b,T,Y_p) + p_{\mu}(n_b, T, Y_\mu) \nonumber \\ &+p_e(n_b, T, Y_e) + p_\gamma(T),\\
    \varepsilon = \varepsilon_b(n_b,T,Y_p) + \varepsilon_{\mu}(n_b, T, Y_\mu) \nonumber \\ &+ \varepsilon_e(n_b, T, Y_e) + \varepsilon_\gamma(T),
\end{align}
where
\begin{eqnarray}
    \varepsilon_\gamma = \frac{8\pi^5}{15}\frac{T^4}{(hc)^3}, \hspace{1cm} p_\gamma = \varepsilon_\gamma/3,
\end{eqnarray}
correspond to the EoS of photons at zero chemical potential and in thermal equilibrium with the matter. Note that we neglect the pressure exerted by trapped neutrinos and their contribution to the matter energy density. See, e.g.,~\cite{Perego:2019adq, Loffredo:2022prq, Pajkos:2025oyf} for investigations concerning the role of trapped neutrinos.

For the purposes of this work, there are two special cases of interest for the muonic EoS: the first one is when $Y_\mu = 0$, which trivially sets $\eta_{\mu^-}^0 = \eta_{\mu^+}^0 = -1/\beta_\mu$. The second one, relevant for the construction of initial data, comes by imposing the neutrinoless $\beta$-equilibrium condition for the reactions
\begin{eqnarray}
    n + \nu_l \leftrightarrow l^- + p.
\end{eqnarray}
In this case, the neutrino chemical potential vanishes and the lepton chemical potentials are given by
\begin{eqnarray}\label{eq:beta-eq}
    &&\mu_e(n_b, T, Y_p) = \mu_\mu(n_b, T, Y_p) = \nonumber \\
    &&\hspace{2cm} \mu_n(n_b, T, Y_p) - \mu_p   (n_b, T, Y_p).
\end{eqnarray}
Setting a constant temperature $T = T_0$, e.g., the lowest tabulated temperature of the baryonic EoS, Eq.~\eqref{eq:ch-neut} is solved for each $n_b$ along with Eq.~\eqref{eq:beta-eq} for $Y_p$. Then, adding the leptons and photons contributions according to Eqs.~\eqref{eq:e-dens},~\eqref{eq:press}, a one-dimensional cold, neutrinoless $\beta$-equilibrated EoS is produced. It is worth pointing out that this EoS is employed only for the construction of initial data, while the full 3-dimensional or 4-dimensional EoSs are employed during the dynamical evolution. Obviously, for simulations without muons, the one-dimensional, neutrinoless $\beta$-equilibrated EoS is obtained by fixing $T = T_0$ and solving
\begin{eqnarray*}
    &&\mu_e(n_b, T_0, Y_p) = \mu_n(n_b, T_0, Y_p) - \mu_p(n_b,T_0,Y_p), \\
    &&Y_e = Y_p.
\end{eqnarray*}

\begin{figure}
    \centering
    \includegraphics[width = 0.95\columnwidth]{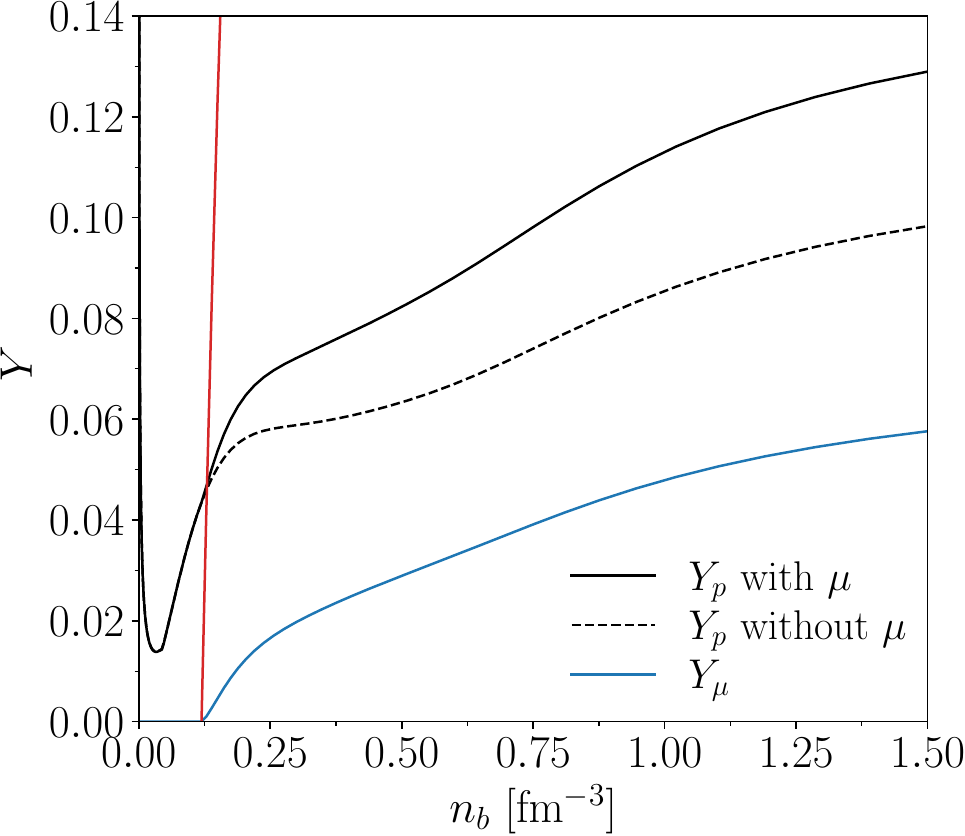}
    \caption{Proton fraction for a cold, $\beta$-equilibrated slice of the SFHo EoS with muons (black thick line), its correspondent muon fraction (blue thick line), and the proton fraction without muons (black dashed line) as a function of the baryonic number density $n_b$. The red vertical line represents the fractional difference $(\mu_e - m_\mu c^2)/(m_\mu c^2)$. For $\mu_e \geq m_\mu c^2$, muons are present.} 
    \label{fig:Yp-beta-eq}
\end{figure}

To illustrate the changes introduced by the presence of muons in the composition of the matter, we depict in Fig.~\ref{fig:Yp-beta-eq} the proton and muon fraction for a cold $T = 0.1~\rm{MeV}$, $\beta$-equilibrated muonic EoS adopting the SFHo baryonic EoS~\cite{Steiner:2012rk}. For small baryon densities $n_b \lesssim 0.125~\rm{fm^{-3}}$, the muonic (thick black line) and electronic (dashed black line) EoSs have the same proton fraction. Once $\mu_e \geq m_\mu c^2$ (which is represented by the red line), muons are present within the matter ($Y_\mu > 0$) and, accordingly, due to the local charge neutrality condition, the proton fraction becomes larger for the muonic EoS by a factor of at most $31\%$ at $n_b = 1.5~\rm{fm^{-3}}$. It should be noted that this leads to a slight decrease of internal energy compared to when muons are absent because of (i) the conversion of electrons into muons and the subsequent loss of electron degeneracy energy by de-occupation of energy levels, which is consistent with the (small) reduction of the electron chemical potential when muons are present, and (ii) the larger proton fraction leads to a loss of the neutron energy contribution to the internal energy. But since muons only appear at moderately high densities, where the baryonic pressure and energy density dominate over the leptonic contributions, the impact of muons in macroscopic properties of a cold NS (e.g., mass, radius and tidal deformability) is negligible.

\begin{figure}[htp!]
    \centering
    \includegraphics[width = 0.96\columnwidth]{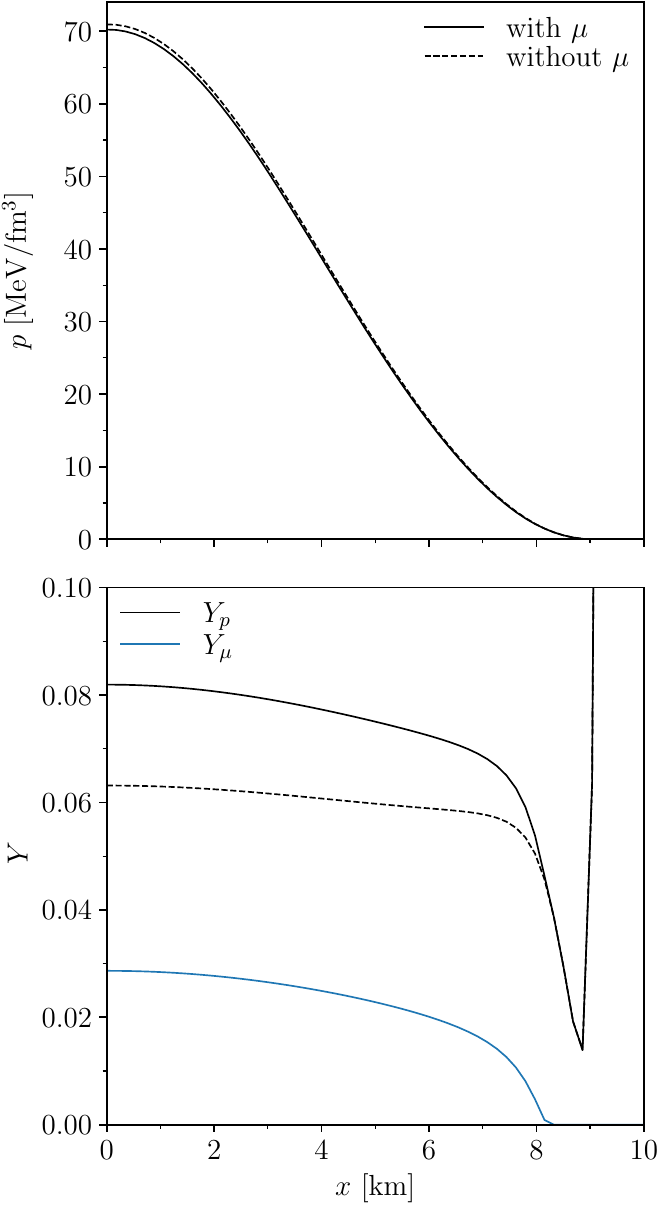}
    \caption{Pressure (upper panel) and particle fractions (lower panel) observed in the initial data with (thick lines) and without (dashed lines) muons.} 
    \label{fig:ID-radial}
\end{figure}

As for the initial data used in our BNS simulations, we present in Fig.~\ref{fig:ID-radial} 1-dimensional profiles along the $x$-axis (in simulation coordinates, where here we adapt $x=0$ to coincide with the center of the NS) of the pressure and particle fractions for the initial data employed in our simulations. As anticipated, for cold, catalyzed matter the pressure (upper panel) is slightly larger around the center of the NS when muons are absent (dashed lines), due to the increased neutron degeneracy pressure in a more neutron-rich medium, indicated by the overall smaller proton fraction in the NS interior (dashed black line in the lower panel). Furthermore, $Y_p$ decreases outward until the interface with the artificial atmosphere adopted in our simulations is reached, which is marked by an increase in $Y_p$ to $\sim 0.46$ (set by the $\beta$-equilibrium condition at the minimum tabulated temperature and density). Finally, for the EoS containing muons, the muon fraction (thick blue line in the lower panel) decreases smoothly toward the edge of the NS, leaving a muonless layer of $\sim 1~{\rm km}$.

In the following, we summarize relevant information regarding the EoS used in this work. For comparison purposes with results from the literature, we adopt the SFHo baseline EoSs. The baryon mass constant is chosen as $m_b = 1.659\times 10^{-24}~\rm{g}$ and the rest-mass density is given by $\rho = m_b n_b$. The range of validity for the EoS parameters are $\rho = [1.695 \times 10^3, 2.489 \times 10^{15}]~\rm{g/cm^3}$, equispaced in log-scale with 30 points per decade, $T = [0.1, 120]~\rm{MeV}$, equispaced in log-scale with 30 points per decade, $Y_p = [0.01, 0.60]$ equispaced in linear scale with stride $0.01$ and $Y_\mu = [1\times 10^{-4}, 1 \times 10^{-1}]$, equispaced in log-scale with 20 points per decade plus $Y_\mu = 0$, for a total of 62 points in the $Y_\mu$ dimension. All necessary thermodynamical information is obtained by means of a quadrilinear interpolation over the aforementioned EoS validity region.
For such a parameterization choice, the 3+1 form of the GRHD equations of~\cite{Font:2008fka} are the same as in~\cite{Galeazzi:2013mia} for the conserved rest-mass density $D$, internal energy density $\tau$ and momentum density $S_i$, but here we evolve $Y_p$ and $Y_\mu$ according to
\begin{eqnarray}
    &&\partial_0(\sqrt{\gamma} D Y_p) + \partial_i[\sqrt{\gamma} D Y_p (\alpha v^i - \beta^i)] = \alpha \sqrt{\gamma} S_{Y_p},\label{eq:prot-adv} \\
    &&\partial_0(\sqrt{\gamma} D Y_\mu) + \partial_i[\sqrt{\gamma} D Y_\mu (\alpha v^i - \beta^i)] = \alpha \sqrt{\gamma} S_{Y_\mu},
\end{eqnarray}
where $\gamma$ is the determinant of the spatial metric, $\alpha$ is the lapse function, $\beta^i$ is the shift vector, $v^i$ is the 3-velocity measured in the Eulerian frame, $S_{Y_p}$ and $ S_{Y_\mu}$ are source terms that we introduce in the next section. The extension of our previous high-resolution shock-capturing (HRSC) scheme~\cite{Gieg:2022} to the present case is straightforward.

\section{Neutrinos Leakage Scheme}
\label{sec:NLS}
\subsection{A Brief Overview}
The NLS~\cite{Ruffert:1995fs, Rosswog:2003rv, OConnor:2009iuz, OConnor:2014sgn, Galeazzi:2013mia, Sekiguchi:2010fh, Deaton:2013sla, Foucart:2014nda, Radice:2016dwd, Perego:2014qda, Perego:2015agy, Palenzuela:2015dqa, Siegel:2017jug, Ardevol-Pulpillo:2018btx, Murguia-Berthier:2021tnt} is a simplified method to account for cooling produced by neutrino irradiation, devised to obtain order-of-magnitude estimates of the role of neutrinos in astrophysical scenarios without resorting to approximate solutions of the Boltzmann equation, such as moments-based schemes~\cite{Shibata:2011kx,Foucart:2015vpa,Foucart:2016rxm,Radice:2021jtw, Anninos_2020, Izquierdo:2022eaz}, Monte-Carlo schemes~\cite{Foucart:2021mcb, Kawaguchi:2022tae, Foucart:2022kon} and lattice-Boltzmann transport schemes~\cite{Weih:2020qyh}. Differently than in previous BNS merger studies, instead of three neutrino species $\{\nu_e, \bar\nu_e, \nu_x\}$, we also consider five neutrino species $\{\nu_e, \bar\nu_e, \nu_\mu, \bar\nu_\mu, \nu_x\}$, where the heavy lepton neutrinos $\nu_x = \{\nu_\tau, \bar\nu_\tau\}$ are grouped as a single species with statistical weight 2. In the following, neutrinos are assumed to be governed by the ultrarelativistic Fermi-Dirac distribution in local thermal and $\beta$-equilibrium with the matter~\cite{Rosswog:2003rv}, i.e., the degeneracy parameters read
\begin{eqnarray}
    &&\eta_{\nu_e} = (\mu_p + \mu_e - \mu_n)/T, \hspace{0.5cm} \eta_{\bar\nu_e} = - \eta_{\nu_e},\label{eq:eta_eq_nue} \\
    &&\eta_{\nu_\mu} = (\mu_p + \mu_\mu - \mu_n)/T, \hspace{0.5cm} \eta_{\bar\nu_\mu} = - \eta_{\nu_\mu}, \label{eq:eta_eq_num}\\
    &&\eta_{\nu_x} = 0,\label{eq:eta_eq_nux}
\end{eqnarray}
where the above chemical potentials include rest-mass.

The NLS prescription adopted in this work assumes that the energy-momentum conservation applied to a matter element is modified according to
\begin{equation}
    \nabla_\nu T_{\rm matter}^{\mu \nu} = -\mathcal{Q} u^\mu,
\end{equation}
where $\nabla_\nu$ is the covariant derivative compatible with the spacetime metric $g_{\mu \nu}$, $T_{\rm matter}^{\mu \nu}$ is the stress-energy tensor of matter, considered here as an ideal fluid, $u^\mu$ is the four-velocity of the matter element and $\mathcal{Q}$ is the total effective energy production rate, given by the sum of effective energy production rates from each neutrino species $Q^{\rm eff}_I, I = \{\nu_e, \bar\nu_e, \nu_\mu, \bar\nu_\mu, \nu_x\}$
\begin{equation}
    \mathcal{Q} = \sum_I Q^{\rm eff}_{I}.
\end{equation}

As stated, the standard NLS only accounts for the cooling of the matter by emission of neutrinos, while absorption and advection of neutrinos are not considered.
On the other hand, since neutrinos are leptons, the set of considered neutrinos-driven reactions consistently modify the lepton family number conservation laws as
\begin{eqnarray}
    \nabla_\nu(\rho Y_e u^\nu) &=& S_{Y_e}, \label{eq:el-cons}\\
    \nabla_\nu(\rho Y_\mu u^\nu) &=& S_{Y_\mu}, \label{eq:m-cons}
\end{eqnarray}
while the baryon number conservation law reads
\begin{equation}
    \nabla_\nu(\rho u^\nu) = 0.\label{eq:bar-cons}
\end{equation}
In face of the above equation, Eqs.~\eqref{eq:el-cons},~\eqref{eq:m-cons} become, respectively
\begin{eqnarray}
    S_{Y_e} = \rho u^\nu \nabla_\nu(Y_e) = \rho \frac{dY_e}{d\tau} \equiv m_b (R^{\rm eff}_{\bar\nu_e} - R^{\rm eff}_{\nu_e}), \\
    S_{Y_\mu} = \rho u^\nu \nabla_\nu(Y_\mu) = \rho \frac{dY_\mu}{d\tau} \equiv m_b (R^{\rm eff}_{\bar\nu_\mu} - R^{\rm eff}_{\nu_\mu}),
\end{eqnarray}
where $d/d\tau$ is the derivative with respect to the proper time of a fluid element. Hence $R^{\rm eff}_{I}$ is interpreted as the effective particle production rate of $I$ in the fluid rest-frame. Finally, applying the local charge neutrality condition Eq.~\eqref{eq:ch-neut}, the source term for $Y_p$ in Eq.~\eqref{eq:prot-adv} reads 
\begin{equation}\label{eq:source-yp}
    S_{Y_p} = S_{Y_e} + S_{Y_\mu} = m_b(R^{\rm eff}_{\bar\nu_e} - R^{\rm eff}_{\nu_e} + R^{\rm eff}_{\bar\nu_\mu} - R^{\rm eff}_{\nu_\mu}).
\end{equation}

Note that in our scheme, $Y_p$ is evolved instead of $Y_e$ via Eq.~\eqref{eq:prot-adv} with source term Eq.~\eqref{eq:source-yp}.

Following~\cite{Ruffert:1995fs,Radice:2018pdn,Siegel:2017jug,Murguia-Berthier:2021tnt}, the effective energy and particle production rates are computed, respectively, according to
\begin{eqnarray}
    Q^{\rm eff}_I &=& Q_I\left(1 + t^{\rm diff}_{I, 1}/ t^{\rm prod}_{I, 1}\right)^{-1}, \label{eq:Qeff}\\
    R^{\rm eff}_I &=& R_I\left(1 + t^{\rm diff}_{I, 0}/ t^{\rm prod}_{I, 0}\right)^{-1},\label{eq:Reff}
\end{eqnarray}
where $Q_I,~R_I$ are the free energy and particle production rates, the production timescales are
\begin{equation}
    t^{\rm prod}_{I, 1} = B_{I, 1}/Q_I,~t^{\rm prod}_{I, 0} = B_{I, 0}/R_I,
\end{equation}
with the neutrino energy density $B_{I, 1}$
\begin{equation}
    B_{I, 1} = g_I \frac{4 \pi}{(hc)^3} T^4 F_3 (\eta_I),
\end{equation}
the neutrino number density $B_{I, 0}$
\begin{equation}
    B_{I, 0} = g_I \frac{4 \pi}{(hc)^3} T^3 F_2 (\eta_I),
\end{equation}
$F_k(\eta_I)$ the ultrarelativistic Fermi integral of order $k$ and the degeneracy factors $g_{\nu_e} = g_{\bar\nu_e} = g_{\nu_\mu} = g_{\bar\nu_\mu} = 1$, $g_{\nu_x} = 2$ ($g_{\nu_x} = 4$) for five (three) neutrinos species. Note that given a set of reactions that produce neutrinos, all the aforementioned quantities may be estimated within our approach by direct interpolation from the EoS since thermal and chemical equilibrium is assumed.

However, the estimation of the diffusion timescale $t^{\rm diff}_{I, 0}$ ($t^{\rm diff}_{I, 1}$) is more involved, since it depends on the local number-averaged opacitiy $\kappa_{I,0}$ (energy-averaged opacity $\kappa_{I, 1}$), and on the non-local optical depth $\tau_{I,0}$ ($\tau_{I,1}$) according to
\begin{eqnarray}\label{eq:t-diff}
    t^{\rm diff}_{I,j} = \frac{\mathcal{D}\tau_{I, j}^2}{c \kappa_{I, j}},~j=\{0,1\},
\end{eqnarray}
with $\mathcal{D} = 6$ chosen following~\cite{OConnor:2009iuz}. For future convenience, we define here the $I$ neutrino-sphere as the surface where $\tau_{I,j} = 1$, which represents the location outside of which neutrinos are effectively decoupled from matter~\cite{Cusinato:2021zin}.

The optical depths are estimated following the iterative procedure of Ref.~\cite{Neilsen:2014hha}, i.e., during the initial timestep, the optical depths are iterated until convergence for all grid points. During the evolution, optical depths are recomputed at each point once per timestep by a single iteration using the optical depths from the previous timestep. An alternative approach, based on the solution of the Eikonal equation for the optical depths may be found in Ref.~\cite{Palenzuela:2022kqk}.

In the following, we present in details the methods employed for the computation of opacities and emission rates for the processes considered in this work, which are summarized in Table~\ref{tab:reactions}.

\begin{table}[t]
    \caption{Weak reactions considered in this work. All the charged-current processes are computed within the elastic approximation. Note that for pair processes and elastic scatterings, neutrinos of all species may participate.}
\begin{tabular}{l||l}
                          & References         \\ \hline \hline
Charged-Current Processes & \multirow{4}{*}{} \\ \hline
$\nu_e + n \leftrightarrow p + e^-$                   &    \cite{Fischer:2020vie} \cite{Rampp:2002bq}\\
$\bar\nu_e + p \leftrightarrow n + e^+$        &    \cite{Fischer:2020vie} \cite{Rampp:2002bq}\\
$\nu_\mu + n \leftrightarrow p + \mu^-$                   &    \cite{Fischer:2020vie} \cite{Rampp:2002bq}\\
$\bar\nu_\mu + p \leftrightarrow n + \mu^+$        &    \cite{Fischer:2020vie} \cite{Rampp:2002bq} \\\hline \hline
Pair Processes         & \multirow{3}{*}{} \\ \hline
$e^- + e^+ \rightarrow \nu + \bar\nu$ &   \cite{Ruffert:1995fs} \cite{Burrows:2006neutrino}\\ 
$\gamma \rightarrow \nu + \bar\nu$ & \cite{Ruffert:1995fs} \cite{Burrows:2006neutrino}       \\ \hline \hline
Elastic Scattering        &           \\  \hline
$\nu + p \rightarrow \nu + p$        &  \cite{Ruffert:1995fs} \cite{Galeazzi:2013mia}           \\
$\nu + n \rightarrow \nu + n$            &    \cite{Ruffert:1995fs} \cite{Galeazzi:2013mia}     \\
$\nu + A \rightarrow \nu + A$    &    \cite{Ardevol-Pulpillo:2018btx}  
\end{tabular}
    \label{tab:reactions}
\end{table}

\subsection{Opacities Computation}
\label{app:A}
A crucial part of modeling neutrino interactions consists of the evaluation of opacities associated with scattering and absorption reactions. As will become clear, a few differences are found between our opacities estimates and the widely adopted prescription for BNS studies, originally due to Ref.~\cite{Ruffert:1995fs}. Instead, we closely follow Refs.~\cite{Fischer:2020vie, Ng:2023syk}.

We begin by considering that the charged-current (CC) absorption processes of Table~\ref{tab:reactions} may be generically represented as
\begin{equation}\label{eq:abs-react}
    \nu + N_1 \rightarrow l + N_2,
\end{equation}
which corresponds to the absorption of a neutrino $\nu$ by the nucleon $N_1$, yielding the lepton $l$ and the nucleon $N_2$, where $N_1, N_2 = \{n, p\}$. 

For simplicity, we restrict to model such reactions by means of the elastic approximation, i.e., neglecting the momentum transferred to nucleons by neutrinos. In this case, the absorption opacity is given by~\cite{Bruenn:1985, Rampp:2002bq, Fischer:2020vie, Ng:2023syk}
\begin{widetext}
\begin{equation}\label{eq:op-def}
    \kappa_I^{\rm abs}(\epsilon) = \frac{\sigma_0 V_{ud}^2}{(m_e c^2)^2}\frac{(1 + 3 g_A^2)}{4}(\epsilon + Q)^2\sqrt{1-\left(\frac{m_l c^2}{\epsilon + Q}\right)^2}[1 - f_l(\epsilon + Q)]\eta_{12},
\end{equation}
\end{widetext}
where $I = \{\nu_e, \bar\nu_e, \nu_\mu, \bar\nu_\mu, \nu_x\}$, $\kappa^{\rm abs}_{\nu_x}(\epsilon) = 0$, $\epsilon$ is the incoming neutrino energy, the reference cross-section
\begin{equation*}
    \sigma_0 = \frac{4G_F^2(m_ec^2)^2}{\pi(\hbar c)^4}\approx 1.761\times 10^{-44}~\rm{cm^2},
\end{equation*}
$G_F = 1.1664\times10^{-5}~{\rm GeV^{-2}}$ is the Fermi constant (in units of $\hbar = c=1$), $V_{ud} = 0.9742$ is the up-down entry of the Cabibbo-Kobayashi-Maskawa matrix~\cite{ParticleDataGroup:2018ovx}, the axial coupling constant is $g_A \approx 1.27$, $E_l = \epsilon + Q$ is the energy of the lepton $l$ and the medium-modified $Q$ value is
\begin{equation}
    Q = m_1^{*}c^2 + U_1 - m_2^* c^2 - U_2,
\end{equation}
where $m_{1/2}^*$ is the effective mass and $U_{1/2}$ is the single-particle vector-interaction potential of $N_{1/2}$, generally provided by the EoS. Otherwise, estimates of $U$ may be obtained following the procedure of~\cite{Martinez-Pinedo:2012eaj}. The lepton distribution function $f_l$ is the Fermi-Dirac function
\begin{equation}
    f_l(\epsilon + Q) = \frac{1}{1+\exp[(\epsilon + Q)/T - \eta_l]},
\end{equation}
and the nucleon phase-space blocking factor $\eta_{12}$ is
\begin{equation}
    \eta_{12} = \frac{n_2 - n_1}{\exp{[(\mu_2 - \mu_1 + Q)/T]}-1},
\end{equation}
where $n_{1/2}$ is the number density of free nucleons and $\mu_{1/2}$ is the nucleon chemical potential (including rest-mass). 

To avoid unphysical behavior of $\eta_{12}$ in the non-degenerate regime, we follow the prescription found in~\cite{Kuroda:2016} and set
\begin{eqnarray}
    \eta_{np} &=& n_n, \\
    \eta_{pn} &=& n_p,
\end{eqnarray}
if $\mu_n - \mu_p - Q < 0.01~{\rm MeV}$.
Note that, although no other corrections are consided, we include in-medium effects in the limited kinematics of the absorption reaction Eq.~\eqref{eq:abs-react} by means of the medium-modified $Q$ factor.

The next step, common to all energy-independent schemes, is to consider the spectral-average of the absorption opacity Eq.~\eqref{eq:op-def}.
To do so, the usual procedure consists in dropping the square root term in Eq.~\eqref{eq:op-def}, which is equivalent to state that the energy of the outcoming lepton $E_l \gg m_l c^2$, i.e., that the produced leptons are ultrarelativistic. This is reasonable for electrons, since in general $Q > m_e c^2$, but for the case of muons, due to their substantially larger rest-mass, such an approximation is not adequate.

Instead, we keep the square-root term, but follow \cite{Ruffert:1995fs, Ardevol-Pulpillo:2018btx} and approximate the lepton phase-space blocking through averaging the energy $\bar{E}_l$ of the produced lepton via reaction Eq.~\eqref{eq:abs-react}
\begin{equation}\label{eq:lep_ph_sp_block}
    [1 - f_l(\epsilon + Q)] \approx \langle 1 - f_l(\bar{E}_l) \rangle = \left\{1 + \exp{[-(\bar{E}_l/T - \eta_l)]}\right\}^{-1}.
\end{equation}
Hence, the spectrally-averaged absorption opacity reads
\begin{eqnarray}\label{eq:sp-avg-op}
    \kappa^{\rm abs}_{I,j} = \frac{1}{B_{I,j}} \langle 1 - f_l(\bar{E}_l) \rangle \frac{\sigma_0 V_{ud}^2}{(m_e c^2)^2}\frac{(1 + 3 g_A^2)}{4} \eta_{12} \mathcal{I}_{I,j}, \nonumber\\
\end{eqnarray}
which is written in terms of the integral
\begin{widetext}
\begin{equation}\label{eq:int-def}
    \mathcal{I}_{I,j}(m_l, Q, T, \eta_I) = \frac{4\pi}{(hc)^3}T^{5+j}\int_{x_{\min}}^{\infty}(x + Q/T)^2\sqrt{1-\left(\frac{m_l c^2}{xT + Q}\right)^2} x^{2+j}f_{I}(x)dx.
\end{equation}
\end{widetext}

In Eq.~\eqref{eq:int-def} the lower integration limit is $x_{\min} = \max[0, (m_l c^2 - Q)/T]$, which ensures that (i) the square-root term is real and (ii) that only neutrinos with energies larger than $m_l c^2 - Q > 0$ are absorbed. Naturally, $f_I(x)$ is the ultrarelativistic Fermi Dirac distribution function describing neutrinos, i.e.,
\begin{equation}
    f_I(x) = \frac{1}{1 + \exp(x - \eta_I)}.
\end{equation}
Before proceeding to the methods employed to perform the integral Eq.~\eqref{eq:int-def}, we are in position of defining the average energy of the produced lepton $\bar{E}_l$ by noting that the average energy of the absorbed neutrinos may be estimated as
\begin{equation}
    \bar{E}_I = \kappa^{\rm abs}_{I, 1}/\kappa^{\rm abs}_{I,0} = T(\mathcal{I}_{I,1}/\mathcal{I}_{I,0}).
\end{equation} 
Thus, energy conservation implies
\begin{equation}\label{eq:E_lep_avg}
    \bar{E}_l = T\frac{\mathcal{I}_{I, 1}}{\mathcal{I}_{I,0}} + Q.
\end{equation}
We verified that such a prescription, along with the lower integration bound $x_{\min}$ defined later ensures that $\bar{E}_l \geq m_l c^2$. It is straightforward to verify that when the square-root term of the integral Eq.~\eqref{eq:int-def} is neglected, one recovers Eq.~(B13) of Ref.~\cite{Ardevol-Pulpillo:2018btx} from Eq.~\eqref{eq:sp-avg-op}. Furthermore, neglecting $Q$, one recovers the widely adopted estimate of Ref.~\cite{Ruffert:1995fs}
\begin{equation*}
    \bar{E}_l = T\frac{F_5(\eta_I)}{F_4(\eta_I)}.
\end{equation*}

So far we have restated the problem of computing opacities as that of evaluating the integral Eq.~\eqref{eq:int-def}. The current, widely adopted procedure of neglecting the square-root term and the $Q$ factor have the clear advantage of reducing the problem to the evaluation of the ultrarelativistic Fermi-Dirac integral, which is easily computed along a simulation given the pair $(T, \eta_I)$ by means of, e.g., the sufficiently accurate formulas of Ref.~\cite{1978A&A....67..185T}. However, as said, such an approach is not justified when applied to muonic weak reactions. 

On the other hand, the numerical integration of Eq.~\eqref{eq:int-def} on-the-fly is computationally intensive, since it may take up to hundreds of function evaluations per integral. Therefore, we resort to a pre-computation of the integrals as to produce, from the EoS as input, a table of spectrally-averaged opacities and emission rates parameterized by $(\rho, T, Y_p, Y_\mu)$, which are then used to compute opacities and emission rates along our simulations by means of quadrilinear interpolations.

Therefore, for each EoS point $(\rho, T, Y_p, Y_\mu)$, the integration of Eq.~\eqref{eq:int-def} is performed by adaptive quadratures up to desired accuracy with the Double Exponential method~\cite{Takahasi1974721} as implemented in Refs.~\cite{MOHANKUMAR200571, 10.5555/1403886}. Naturally, modifications concerning the kernel of the integral and the lower integration boundary are in order, since the original method is devised to integrate moments of the Fermi-Dirac distribution from $x_{\min} = 0$, which can be handled by simple variable transformations.

More specifically, we first distinguish two cases:
\begin{itemize}
    \item[(i)] If $m_l c^2 - Q \leq 0$, we integrate Eq.~\eqref{eq:int-def} with $x_{\min} = 0$, since neutrinos with all energies may participate of the reaction. The $I$ degeneracy $\bar\eta_I = \eta_I$ and distribution function $\bar{f}_I(x) = f_I(x)$ remain unchanged,\\
    \item[(ii)] If $m_l c^2 - Q > 0$, we make $\epsilon + Q = E + m_l c^2$ and $x = E/T$. The $I$ degeneracy is, thus, re-scaled by the transformation such that
\end{itemize}
    \begin{eqnarray}
        &&\tilde{\eta}_I = \eta_I - \frac{(m_l c^2 - Q)}{T}, \\
        &&\tilde{f}_I(x) = \frac{1}{1+ \exp{(x - \tilde{\eta}_I)}},
    \end{eqnarray}
    and the integral Eq.~\eqref{eq:int-def} becomes
    \begin{widetext}
        \begin{equation}\label{eq:int-tilde-def}
        \tilde{\mathcal{I}}_{I,j} = \frac{4\pi}{(hc)^3}T^{5+j}\int_{0}^{\infty}(x + m_l c^2/T)^2\sqrt{1-\left(\frac{m_l c^2}{xT + m_l c^2}\right)^2} \left(x + \frac{m_l c^2-Q}{T} \right)^{2+j}\tilde{f}_{I}(x)dx,
    \end{equation}
    \end{widetext}
where we omitted the dependencies of $\tilde{\mathcal{I}}_{I, j}$ for a shorter notation.

One last limit has to be considered when computing Eq.~\eqref{eq:sp-avg-op}: the black-body function $B_{I,j}$ may be evaluated to zero, which generally occurs for very negative neutrino degeneracies, although the ratio $\mathcal{I}_{I,j}/B_{I,j}$ is finite. Thus, in order to circumvent such a possible issue, we first compute the black-body functions $B_{I,0},~B_{I,1}$. If one of those functions evaluate to zero, we make
\begin{equation}
    B_{I,j} = \frac{4\pi}{(hc)^3} T^{3+j}\exp(\eta_I)(2+j)!,
\end{equation}
which comes from expanding $f_I(x) \approx \exp(\eta_I)\exp(-x)$ for $\exp(\eta_I) \ll 1$.

On the other hand, when $\eta_I \leq -100$ (for $m_l c^2 - Q \leq 0$) or $\bar\eta_I \leq -100$ (for $m_l c^2 - Q > 0$), we proceed to similar expansions
\begin{eqnarray}
    &&f_I(x) \approx \exp(\eta_I)\exp(-x), \\
    &&\bar{f}_I(x) \approx \exp(\bar\eta_I)\exp(-x),
\end{eqnarray}
and carry out the integrations Eq.~\eqref{eq:int-def} or Eq.~\eqref{eq:int-tilde-def} with a 64 points Gauss-Laguerre quadrature. By doing so, we have explicitly factored out the $\exp(\eta_I)$ term that may drive $B_{I,j} \rightarrow 0$, thus allowing the computation of finite ratios $\mathcal{I}_{I,j}/B_{I,j}$.

Finally, for the elastic scattering of neutrinos $\nu$ in free nucleons $N$ and heavy nuclei $A$
\begin{eqnarray}
    \nu + N &\rightarrow& \nu + N, \\
    \nu + A &\rightarrow& \nu + A,
\end{eqnarray}
we compute the respective scattering opacities $\kappa^{\rm scatt}_{I, j}(N)$ according to~\cite{Galeazzi:2013mia} and $\kappa^{\rm scatt}_{I, j}(A)$ according to~\cite{Ardevol-Pulpillo:2018btx}. Hence, the opacities used in Eq.~\eqref{eq:t-diff} and for the computation of optical depths is simply the sum of the opacities over all processes, i.e.,
\begin{eqnarray}
    {\small \kappa_{I, j} = \kappa_{I, j}^{\rm abs} + \kappa^{\rm scatt}_{I, j}(n) + \kappa^{\rm scatt}_{I, j}(p)+ \kappa^{\rm scatt}_{I, j}(A),}
\end{eqnarray}
with $\kappa_{\nu_x, j}^{\rm abs} = 0$.\\

\subsection{Emission Rates Computation}
\label{app:B}

For the emission via charged-current processes we consider the inverse of the reaction presented in Eq.~\eqref{eq:abs-react}. Following~\cite{Bruenn:1985}, detailed-balance sets the spectrally-averaged emission rates as
    \begin{eqnarray}\label{eq:sp-emiss-def}
    Q^{\rm CC}_{I,j} = \langle 1 - f_I(\bar{E}_I) \rangle \frac{\sigma_0 V_{ud}^2 c}{(m_e c^2)^2}\frac{(1 + 3 g_A^2)}{4} \eta_{21} \mathcal{I}^*_{I,j}, \nonumber \\
\end{eqnarray}
where the integral $\mathcal{I}^*_{I,j}$ reads
\begin{widetext}
\begin{equation}\label{eq:int-emiss-def}
    \mathcal{I}^*_{I,j}(m_l, Q, T, \eta_l) = \frac{4\pi}{(hc)^3}T^{5+j}\int_{x_{\min}}^{\infty}(x + Q/T)^2\sqrt{1-\left(\frac{m_l c^2}{xT + Q}\right)^2} x^{2+j}f_{l}(x+Q/T)dx.
\end{equation}
\end{widetext}
For easy of notation and consistency with the text, we note that $Q^{\rm CC}_{I,0} = R^{\rm CC}_I$, $Q^{\rm CC}_{I,1} = Q^{\rm CC}_I$ and $Q^{\rm CC}_{\nu_x, 1} = Q^{\rm CC}_{\nu_x, 0} = 0$.

In this case the neutrinos produce the phase-space blocking, thus in complete analogy to Eq.~\eqref{eq:lep_ph_sp_block} we define
\begin{equation}\label{eq:neu_ph_sp_block}
    \langle 1 - f_I(\bar{E}_I) \rangle = \left\{1 + \exp{[-(\bar{E}_I/T - \eta_I)]}\right\}^{-1},
\end{equation}
such that the average energy of the produced neutrino $\bar{E}_I$ is given by
\begin{equation}
    \bar{E}_I = \max \left[ 0, T \frac{\mathcal{I}^*_{I,1}}{\mathcal{I}^*_{I,0}} - Q\right].
\end{equation}

Similarly to the calculation of absorption opacities, we distinguish two cases:
\begin{itemize}
    \item[(i)] If $m_l c^2 - Q \leq 0$, we set $x_{\min} = 0$ and compute the integral Eq.~\eqref{eq:int-emiss-def} with the modified lepton distribution function $f^*_l(x)$
    \begin{eqnarray}
        &&f^*_l(x) = f_l(x + Q/T) = \frac{1}{1 + \exp{(x - \eta^*_l)}}, \\
        &&\eta^*_l = \eta_l - \frac{Q}{T},
    \end{eqnarray}
    which is the same integral as Eq.~\eqref{eq:int-def}, up to a substitution $f_I(x) \rightarrow f^*_l(x)$.
    \item[(ii)] If $m_l c^2 - Q > 0$, we make again $\epsilon + Q = E + m_l c^2$, $x = E/T$, which transforms the lepton distribution function and the lepton chemical potential, respectively, as
    \begin{eqnarray}
        &&\tilde{f}_l^*(x) = \frac{1}{1 + \exp{(x - \bar{\eta}^*_l)}}, \\
        &&\bar{\eta}^*_l = \eta_l - \frac{m_l c^2}{T}.
    \end{eqnarray}
\end{itemize}
The resulting integral, then, is the same as Eq.~\eqref{eq:int-tilde-def}, up to a substitution $\tilde{f}_I(x) \rightarrow \tilde{f}^*_l(x)$.

As per the pair processes, we follow the expressions of Ref.~\cite{Ruffert:1995fs} for the electron-positron pair annihilation~($e^- e^+$) and transversal plasmon decay~($\gamma$) with a few adaptations, namely: 
\begin{itemize}
    \item[(i)] For $\nu_x$ we divide their Eqs.~(B10),~(B12) by 2 to account for our statistical weight 2 instead of 4.
    \item[(ii)]  For $\nu_\mu$ and $\bar\nu_\mu$ produced via $e^-e^+$, we use their Eq.~(B8), changing the term $(C_1 + C_2)_{\nu_e\bar\nu_e} \rightarrow (C_1 + C_2)_{\nu_x\bar\nu_x}$ and employ the degeneracies Eq.~\eqref{eq:eta_eq_num} to compute the corresponding blocking factors in their Eq.~(B9).
    \item[(iii)] Analogous adaptions were made in their Eqs.~(B11),~(B13) for the production of $\nu_\mu$ and $\bar\nu_\mu$ via $\gamma$.
\end{itemize}

Then the free production rates in Eqs.~\eqref{eq:Qeff},~\eqref{eq:Reff} are given by the sum of production rates over the charged-current and pair processes.

\subsection{Limitations of the Elastic Approximation}{\label{sec:lim-el-ap}}

Although in this work we employ the elastic approximation for the computation of semi-leptonic neutrino opacities, in line with most of the work developed so far in the context of BNS merger simulations, it is important to point out that there is room for improvements in the modeling of weak neutrino interactions, for instance, the adoption of relativistic kinematics for nucleons~\cite{Roberts:2016mwj}, the inclusion of nuclear correlations in dense medium~\cite{Oertel:2020pcg}, or the full kinematics treatment~\cite{Guo:2020tgx}. In this Section, we focus on comparing the opacities of charged-current processes obtained by Eq.~\eqref{eq:op-def} (employed in this work) with the full kinematics approach of Ref.~\cite{Guo:2020tgx}, in which weak magnetism, pseudoscalar and nuclear form factor are considered in a relativistic framework. Such a comparison was carried out in detail in Refs.~\cite{Fischer:2018kdt, Guo:2020tgx, Ng:2023syk} for conditions relevant to supernova matter, but here we focus on conditions found in BNS merger remnants, summarized in Table~\ref{tab:td-points}, where point A corresponds to the remnant core, point B to the hot, high-density interface between the core and the disk, and point C to the colder, low-density disk.

\begin{figure*}[ht!]
    \centering
    \includegraphics[width=\linewidth]{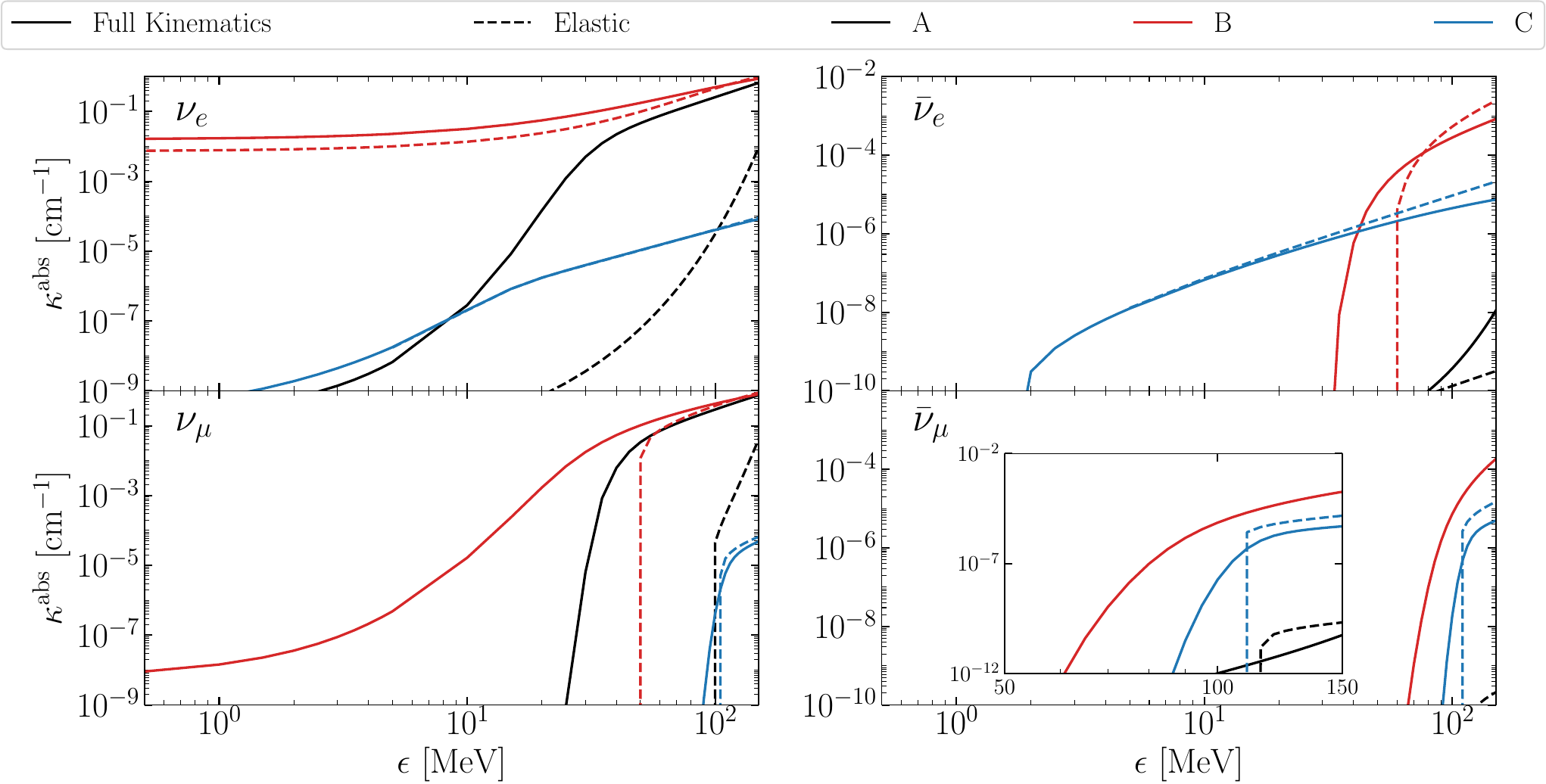}
    \caption{Spectral absorption opacities for charged-current processes considered in this work in the full kinematics approach (solid lines) and in the elastic approximation (dashed lines) at relevant thermodynamical conditions A (black), B (red) and C (blue), representing the remnant core, hot interface and disk, respectively.}
    \label{fig:spec-opac}
\end{figure*}

In Fig.~\ref{fig:spec-opac} we show the spectral absorption opacity for each neutrino species in the full kinematics approach (solid lines) and in the elastic approximation (dashed lines) for each of the aforementioned thermodynamical points. First we note that the effects of the full kinematics treatment is more pronounced at high densities, where the momentum transfer becomes important (especially for $\nu_\mu,~\bar\nu_\mu$ due to the high muon rest-mass), as seen by the substantial opacity enhancement at points A and B across all neutrino species. The differences in opacity at point C (blue lines) are small for $\nu_e,~\nu_\mu$, but sizable for $\bar\nu_e$ ($\bar\nu_\mu$) at $\epsilon \geq 45~{\rm MeV}$ ($\epsilon \geq 90~{\rm MeV}$) with up to an order of magnitude higher opacities in the elastic approximation. Muon (anti)neutrinos are expected to freely stream in this region, given that their average energy in equilibrium ($\langle \epsilon_{\nu_\mu,\bar\nu_\mu}\rangle\ =  TF_3(\eta_{\nu_\mu, \bar\nu_\mu})/F_2(\eta_{\nu_\mu, \bar\nu_\mu}) \approx 10~{\rm MeV}$) lies in the suppressed region.

In the remnant core (point A, black lines) we observe that $\nu_e$ absorption is greatly enhanced in the full kinematics approach at all neutrino energies, between several orders of magnitude for $\epsilon \leq 20~{\rm MeV}$ and two orders of magnitude at $\epsilon = 150~{\rm MeV}$. The effect is more dramatic in the case of $\nu_\mu$, where the suppression threshold is reduced from $\epsilon = 98~{\rm MeV}$ to $\epsilon = 32~{\rm MeV}$. Hence, the full kinematics approach should impact the neutrino trapping and thermalization in the core and consequently the conditions under which neutrinos of different energies decouple from matter~\cite{Endrizzi:2019trv}. Similar features are observed for $\bar\nu_e$ and $\bar\nu_\mu$, but given their small absorption opacities in this regime, it is expected that scattering will dominate the (fast) thermalization of antineutrinos and no significant effect should arise from the full kinematics prescription.

In the hot core-disk interface (point B, red lines) we note that the absorption opacity for $\nu_e$ does not exhibit suppression at low energies and remains larger in the full kinematics approach by a factor of $1 - 2.5$ for energies below $100~{\rm MeV}$, while for $\bar\nu_e$ the energy threshold decreases from $\sim 55~{\rm MeV}$ to $\sim 32~{\rm MeV}$. The hard cut-off introduced by the elastic approximation at $\sim 50~{\rm MeV}$ ($\sim 160~{\rm MeV}$) for $\nu_\mu$ ($\bar\nu_\mu$) is absent (relaxed to $\sim 61~{\rm MeV}$) with the full kinematics treatment. Therefore, we expect that the substantial opacity increase across species could extend the neutrinospheres further out of the remnant, leading to less effective neutrino cooling and less intense leptonization or deleptonization of the remnant, depending on the new positions of the neutrinospheres.

Here we make a remark regarding the possible impacts of electron neutrino capture on nuclei $A$ with $N$ neutrons and $Z$ protons,
\begin{equation}
    \nu_e + A'(N+1, Z-1) \rightarrow A(N,Z) + e^-,
\end{equation}
which was not considered in the present work. Using the NuLib library~\cite{Sullivan:2015kva}, we computed the associated opacity based on weak rate tables of Refs.~\cite{Suzuki:2017vzi, Oda:1994bwl, Langanke:2003ii, Langanke:2000ii}. For the conditions A-C in Table~\ref{tab:td-points}, the opacity is negligible due to the small fraction of nuclei. In fact, this reaction has non-negligible opacity at ejecta conditions, i.e., densities $\rho \lesssim 10^{10}~{\rm g/cm^3}$ and temperatures $T \lesssim 1~{\rm MeV}$, but only at relatively large proton fraction $Y_p \gtrsim 0.4$. Due to the adoption of a leakage scheme, the later condition is never met in our simulations, as a full treatment of absorption seems to be necessary to produce highly protonized ejecta~\cite{Zappa:2022rpd}. Hence, in scenarios where low density material is expected to become very proton-rich, this reaction should not be ignored, since it may dominate electron-neutrino capture on free neutrons.

As a last point, in the following we investigate the impact of neglecting weak magnetism/recoil corrections (WM) in the form of multiplicative, energy-dependent factors, as proposed by Ref.~\cite{Horowitz:2002weak}. To do so, we consider the same representative points of Table~\ref{tab:td-points} and compute the energy-weighted opacities for the various neutrino species $I$
\begin{equation}
    \bar\kappa_I(\epsilon) = \frac{\epsilon^3 f_I(\epsilon)}{T^4 F_3(\eta_I)}\kappa_{I}({\epsilon}),
\end{equation}
representing the quantity to be integrated in $\epsilon$ (up to average blocking factors) to produce the gray opacities employed in this work.

\begin{table*}[ht!]
    \caption{Thermodynamic state of representative points extracted from our simulations. From left to right the columns represent the point label, temperature, rest-mass density, proton fraction, muon fraction, neutron chemical potential, proton chemical potential, neutron-proton single particle interaction potential difference, neutron effective mass, proton effective mass, electron chemical potential and muon chemical potential.}
\begin{tabularx}{\textwidth}{XXXXXXXXXXXX}
\hline \hline
Point        &   $T$ & $\rho$ & $Y_p$ & $Y_\mu$ & $\mu_n$ & $\mu_p$ & $U_n - U_p$  & $m^*_n$ & $m^*_p$ & $\mu_e$ & $\mu_\mu$    \\ 
 & [MeV] & [g/cm$^3$] & & & [MeV] & [MeV] & [MeV] & [MeV] & [MeV] & [MeV] & [MeV] \\ \hline
A & $10.0$ & $1.2\times10^{15}$ & $0.08$ & $0.025$ & $1413$ & $1192$ & $7.6$ & $349$ & $ 347$ & $209$ & $192$ \\ \hline
 B & $30.0$ & $5.0\times10^{14}$ & $0.06$ & $0.015$ & $1020$ & $829$ & $54.8$ & $590$ & $ 588$ & $125$ & $112$ \\ \hline
  C & $3.0$ & $1.0\times10^{10}$ & $0.23$ & $0$ & $927$ & $922$ & $0.02$ & $939$ & $ 938$ & $12.6$ & $0$ \\ \hline \hline
\end{tabularx}
    \label{tab:td-points}
\end{table*}

\begin{figure*}[ht!]
    \centering
    \includegraphics[width=\linewidth]{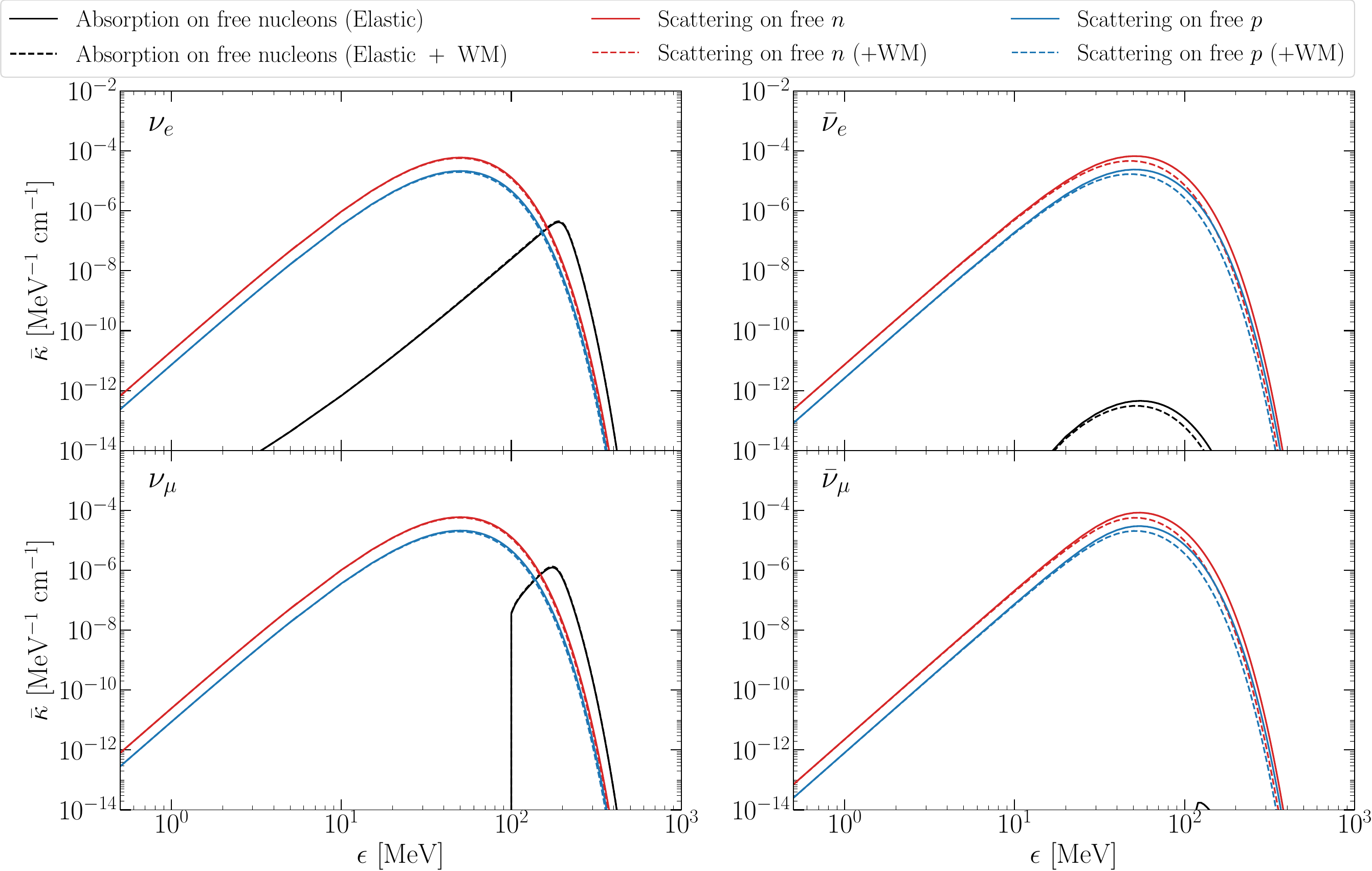}
    \caption{Energy-weighted opacities for the weak reactions considered in this work in the elastic approximation (solid lines) and with weak magnetism/recoil corrections (dashed lines) at thermodynamical point A for electron neutrinos (upper left panel), electron anti-neutrinos (upper right panel), muon neutrinos (lower left panel) and muon anti-neutrinos (lower right panel). The lines represent absorption on free nucleons (black), scattering on free neutrons (red), and scattering on free protons (blue).}
    \label{fig:wm-op-core}
\end{figure*}

For the thermodynamical point A (core), we present in Fig.~\ref{fig:wm-op-core} the energy-weighted opacities. As anticipated, WM corrections are negligible for the reactions involving neutrinos (upper and lower left panels), but reduce opacities at higher neutrino energies for anti-neutrinos (upper and lower right panels). In particular we note that under this thermodynamical condition, neutrinos are expected to be in thermal and weak equilibrium with the fluid. Given that in a gray scheme the integral of the depicted quantity is the relevant quantity to be coupled to the fluid evolution, and that the most noticeable effect is observed for iso-energetic elastic scattering on free nucleons for anti-neutrinos, WM contributes in reducing $\bar\nu_e$ and $\bar\nu_\mu$ opacities by a maximum $\sim 35\%$.

\begin{figure*}[ht!]
    \centering
    \includegraphics[width=\linewidth]{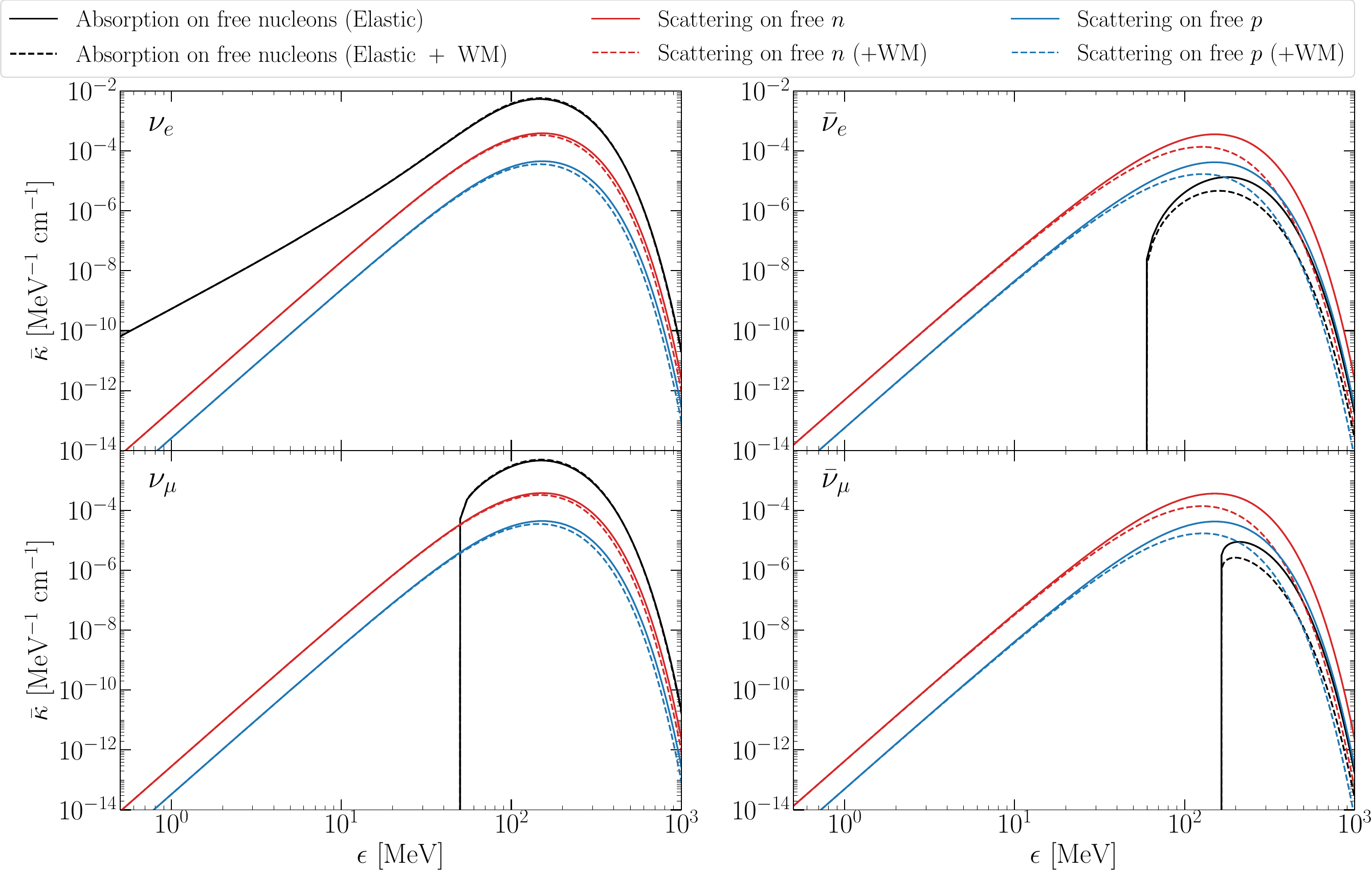}
    \caption{Same as Fig.~\ref{fig:wm-op-core}, but for thermodynamical point B, representing the hot core-disk interface.}
    \label{fig:wm-op-hot}
\end{figure*}

In Fig.~\ref{fig:wm-op-hot} we repeat the analysis for thermodynamical point B (hot core-disk interface). Similarly, we note that WM corrections are small for neutrinos (upper and lower left panels) and most perceivable for iso-energetic elastic scattering on free nucleons. However, CC opacities dominate the integral by $1-2$ orders of magnitude compared to scattering opacities, while WM contributes with a few percent deviation. There we note that WM corrections are once again small for the energy-integrated opacities of neutrinos. On the other hand, WM corrections for anti-neutrinos (upper and lower right panels) noticeably decrease opacities at higher energies, but the weighting against the equilibrium distribution function for the computation of energy-integrated opacities limits this effect, as the anti-neutrino occupation number decreases. In this case, we see an overall reduction of $\sim 70\%$ on energy-integrated opacities when WM corrections are included.

Hence, we have seen that WM corrections are responsible for significantly reducing opacities for anti-neutrinos across semi-leptonic CC and iso-energetic neutrino-nucleon elastic scattering reactions, in particular for thermodynamical conditions where neutrinos should be trapped, thus where a slow diffusion ($t^{\rm diff} \gg t^{\rm prod}$) of thermalized neutrinos dominates the effective production rates [Eqs.~\eqref{eq:Qeff}-\eqref{eq:Reff}]. In this case, one expects very modest contributions of neutrino source terms to the fluid evolution.

In a leakage scheme, regions that are more strongly affected by neutrinos are usually located in the disk and polar cap, at optical depths from order of unit to order of tens, typically corresponding to $\rho = 10^{10}-10^{12}~{\rm g/cm^3}$~\cite{Endrizzi:2019trv}, where neutrinos decouple from matter and more energetic free emission contributes more to the effective rates. Thus, most of the impact of WM corrections in our simulations should come from thermodynamical conditions better represented by point C of Table~\ref{tab:td-points} (outer disk), which coincides with matter conditions approximately at the $\bar\nu_e$ and $\bar\nu_\mu$ neutrinospheres, and points within the neutrinosphere with temperature $T \sim 1-10~{\rm MeV}$ and densities in the aforementioned range.

\begin{figure*}[ht!]
    \centering
    \includegraphics[width=\linewidth]{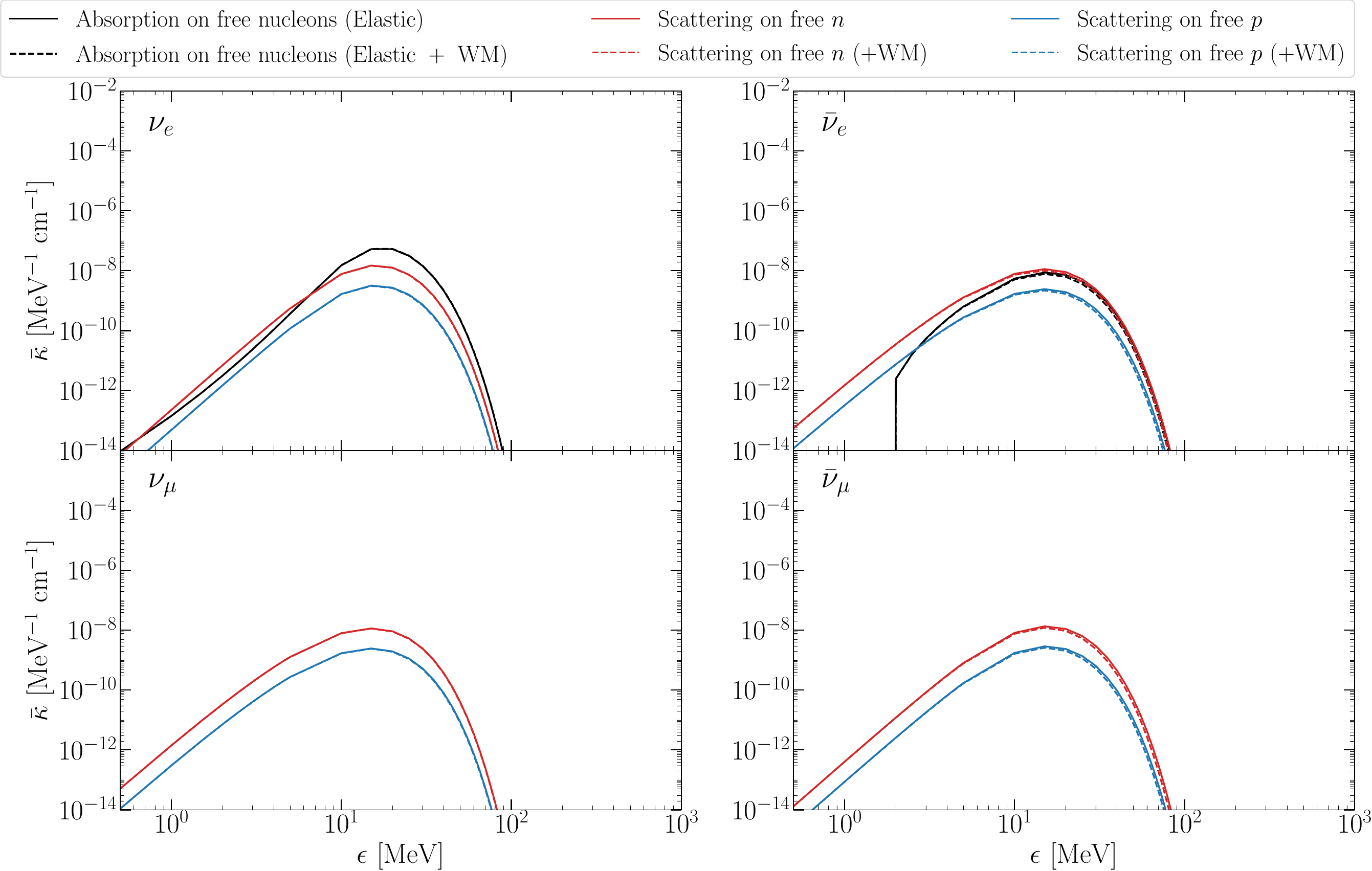}
    \caption{Same as Fig.~\ref{fig:wm-op-core}, but for thermodynamical point C, representing the outer disk.}
    \label{fig:wm-op-disk}
\end{figure*}

From Fig.~\ref{fig:wm-op-disk}, where outer disk conditions are represented, one notices negligible opacity deviations for neutrinos (on the percent level), and small deviations (on the few percent level) for anti-neutrinos. Considering that (i) the optical depth at a point is determined by the line integral of the opacity along a stationary path connecting the grid (or refinement level) boundary to said point, and (ii) that WM corrections become less important away from the remnant, the positions of the neutrinospheres of all species should remain almost unaltered.

Therefore, one expects quantitative changes on simulation results by employing WM corrections with a leakage scheme prescription to be restricted to the region between the outer disk and the hot core-disk interface. In fact, for one such intermediate point (not shown), we observe reductions of $\sim 20 - 25\%$ in the scattering and CC opacities of anti-neutrinos, suggesting impacts of the same order to the electronization of matter.

In summary, the matter dynamics within the core and the hot core-disk layer would be very slightly impacted, due to the high optical depths found in such regions. Further away from the outer disk and on the ejecta, the impacts should also be negligible, as density and temperature are significantly smaller.

It should be noted, however, that most of the differences expected to arise from the adoption of various treatments, either in the form of the full kinematics approach for charged-current processes, or WM/recoil corrections to the elastic rates, would be better discernible with the adoption of a more advanced treatment for neutrinos transport, such as a moments scheme, given that neutrinos capture processes are the most affected by such corrections.
With those limitations in mind, we present in the following the results obtained with the neutrinos leakage scheme and the elastic approximation for the charged-current opacities, suited for a first assessment of the impacts of muons and muonic weak reactions in BNS mergers.

\section{Methods and Setups}
\label{sec:Methods}
In this work we performed four equal-mass, irrotational, BNS merger simulations. For our simulations we employed the SFHo baseline EoS. For comparison purposes, we ran one simulation with electrons and positrons only, which corresponds to a three-dimensional EoS, incorporating the usual three neutrino species $\{\nu_e, \bar\nu_e, \nu_x\}$, $\nu_x = \nu_\mu, \bar\nu_\mu, \nu_\tau, \bar\nu_\tau$, with $g_{\nu_x} = 4$. This setup is referred to as SFHo$\_$3D. The other three setups were simulated with the full four-dimensional EoS. In order to single out the role of muons-driven reactions, one of the four-dimensional EoS setup was simulated with the same aforementioned three neutrino species, hence named SFHo$\_$4D$\_$3, while the remaining two were simulated with five neutrino species $\{\nu_e, \bar\nu_e, \nu_\mu, \bar\nu_\mu, \nu_x\}$, $\nu_x = \nu_\tau,\bar\nu_\tau$, $g_{\nu_x} = 2$, identified as SFHo$\_$4D$\_$5, with same spatial grid resolution as the previously described runs, and SFHo$\_$4D$\_$5$\_$High, with higher spatial grid resolution. All systems have total gravitational mass $M = 2.70~M_\odot$ and are initially governed by cold $T = 0.1~{\rm MeV}$, neutrinoless $\beta$-equilibrated EoSs. The initial data was produced with the SGRID code~\cite{Tichy:2009yr, Tichy:2012rp, Tichy:2019ouu}, adapted to the use of one-dimensional tabulated EoSs as input. One caveat is that finite-temperature EoSs hardly reproduce the limit $(p,~\varepsilon) \rightarrow 0$ for $n_b \rightarrow 0$, which is needed for the proper imposition of boundary conditions. Instead, there is typically a (small) critical density $n_b^*$ for which $de/dn_b|_{n_b^*} = 0$ and such that $de/dn_b > 0$ for $n_b < n_b^*$~\cite{Alford:2022bpp}. To ensure the desired behavior at densities below the critical one, we assume that in this region the EoS is described by a cold polytrope.

The dynamical evolution of the spacetime and matter was performed with the BAM code, which received the updates described in Sec.~\ref{sec:EoS} and in Sec.~\ref{sec:NLS}. The numerical domain consists in a hierarchy of $7$ nested Cartesian grids (referred to as levels and indexed by $L \geq 0$) with $\Delta x_{L}/\Delta x_{L+1} = 2$ grid spacing refinement. For $L \geq 3$ the levels move following the motion of the stars.

The finest level $(L = 6)$ has grid spacing $\Delta x_6 = 178~{\rm m}$ and $\Delta x_6 = 142~{\rm m}$ for the high resolution run. The space is discretized with fourth-order finite differencing, while all fields evolve in time using the method of lines with an explicit fourth-order Runge-Kutta integrator and the Berger-Oliger time stepper~\cite{Berger:1984}. Geometry quantities are evolved with the Z4c formulation of Einstein's equations~\cite{Bernuzzi:2009ex,Hilditch:2012fp} along with the moving punctures method~\cite{Campanelli:2005dd, Bruegmann:2006ulg}. Gauge fields are evolved adopting the $1 +\log$ slicing~\cite{Bona:1994dr} and the Gamma-driver conditions~\cite{Alcubierre:2002kk}. The HRSC scheme adopted for the evolution of matter fields employs the WENOZ reconstruction~\cite{Borges:2008} and the HLL Riemann solver~\cite{Harten:1983, Toro:1999} for the computation of inter-cell fluxes. Finally, a conservative adaptative mesh refinement strategy is used to ensure mass conservation across refinement levels~\cite{Dietrich:2015iva}. The low density regions outside of the stars are treated with an artificial atmosphere prescription according to which matter elements are static and the thermodynamical properties follow from a cold and $\beta$-equilibrated slice of the EoS.

\section{Merger and Post-Merger Dynamics}
\label{sec:M-PM-Dyn}
\subsection{Matter Evolution}
All simulations begin with a coordinate distance between stars $\approx 41.4~{\rm km}$, merging after $\sim 4$ orbits. As expected, the presence of muons and muon-driven neutrino reactions does not affect the orbital dynamics during the inspiral.

\begin{figure}[ht!]
    \begin{subfigure}
    \centering
    \includegraphics[width=\linewidth]{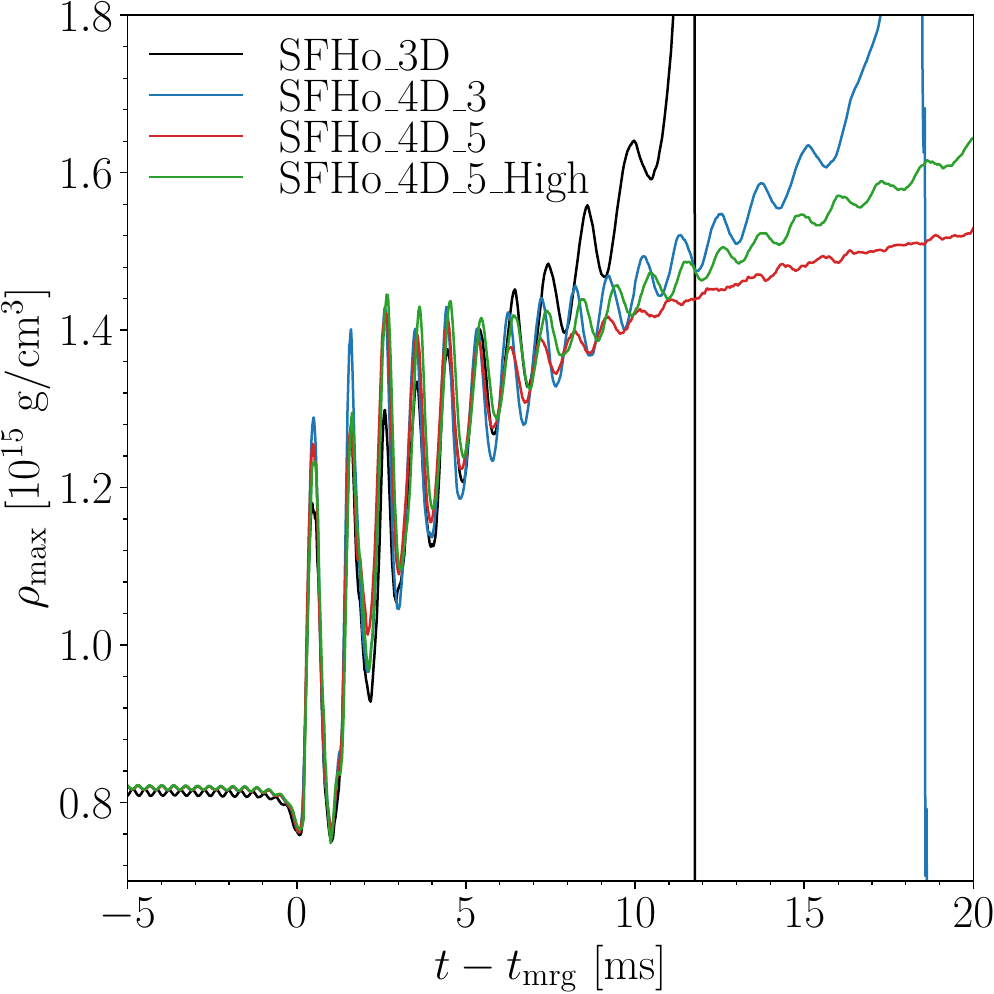}
    \end{subfigure}
    \caption{Maximum rest-mass density evolution for our simulated setups. The merger instant $t_{\rm mrg}$ is defined as the instant in which the amplitude of the dominant $(2,2)$ mode of the GW strain is maximum.}
    \label{fig:rho_max}
\end{figure}

In Fig.~\ref{fig:rho_max} we depict the time evolution of the maximum rest-mass density $\rho_{\rm max}$ for our simulated setups. We begin noting that the muonic EoSs exhibit a slightly larger maximum rest-mass density before the merger, which is a consequence of the slightly smaller internal energy at same density introduced by the inclusion of muons in the cold, $\beta$-equilibrated EoSs employed for the construction of the ID. 

In good agreement with the results from Ref.~\cite{Radice:2018pdn}, our SFHo$\_$3D simulation collapsed to a black-hole in $\sim 12~{\rm ms}$ after the merger which, despite important differences between the hydrodynamics and NLS implementations, is reassuring. Next, we note that the SFHo$\_$4D$\_$3 run has its collapse delayed by $\sim 7~{\rm ms}$ compared to SFHo$\_$3D, which suggests a stabilizing role of the muons in the densest portions of the remnant with a noticeable damping of $\rho_{\rm max}$ oscillations until the collapse.
Furthermore, the remnant of the 5-species run SFHo$\_$4D$\_$5 experiences a stronger damping of the oscillations and does not collapse within our simulation time span. It is worth pointing out that gravitational collapses are rather sensitive to grid resolution, thus the observation of a longer-lived remnant in SFHo$\_$4D$\_$5$\_$High, albeit exhibiting weaker damping as the oscillations are sustained for longer, suggests that the stabilization is robust.

For a more complete description of the reported stabilization, it is useful to introduce the reduced gravitational binding energy $E_b$ and reduced angular momentum $j_{\rm red}$, defined as~\cite{Bernuzzi:2012ci, Zappa:2022rpd}
\begin{eqnarray}
    E_b &=& \frac{1}{\mu M}\left(M_{\rm ADM} - E_{\rm GW} - M\right), \\
    j_{\rm red} &=& \frac{1}{\mu M^2}\left(J_{\rm ADM} - J_{\rm GW} \right),
\end{eqnarray}
where $M_{ADM}$ ($J_{\rm ADM}$) is the ADM mass (angular momentum) of the system at the initial time slice, $M = M_1 + M_2$ is the total gravitational mass , $E_{\rm GW}$ ($J_{\rm GW}$) is the energy (angular momentum) radiated by GWs (see Ref.~\cite{Bernuzzi:2012ci} for the definitions of $E_{\rm GW}$ and $J_{\rm GW}$) and $\mu = M_1 M_2/(M_1 + M_2)^2$ is the symmetric mass ratio. For our simulations, $M_{\rm ADM} = 2.672~M_\odot$, $M = 2.7~M_\odot$, $J_{\rm ADM} = 7.064~M_\odot^2$ and $\mu = 1/4$. Physically, $E_b$ characterizes the compactness of the system, while $j_{\rm red}$ represents its angular momentum.

\begin{figure}[ht!]
    \begin{subfigure}
    \centering
    \includegraphics[width=0.9\linewidth]{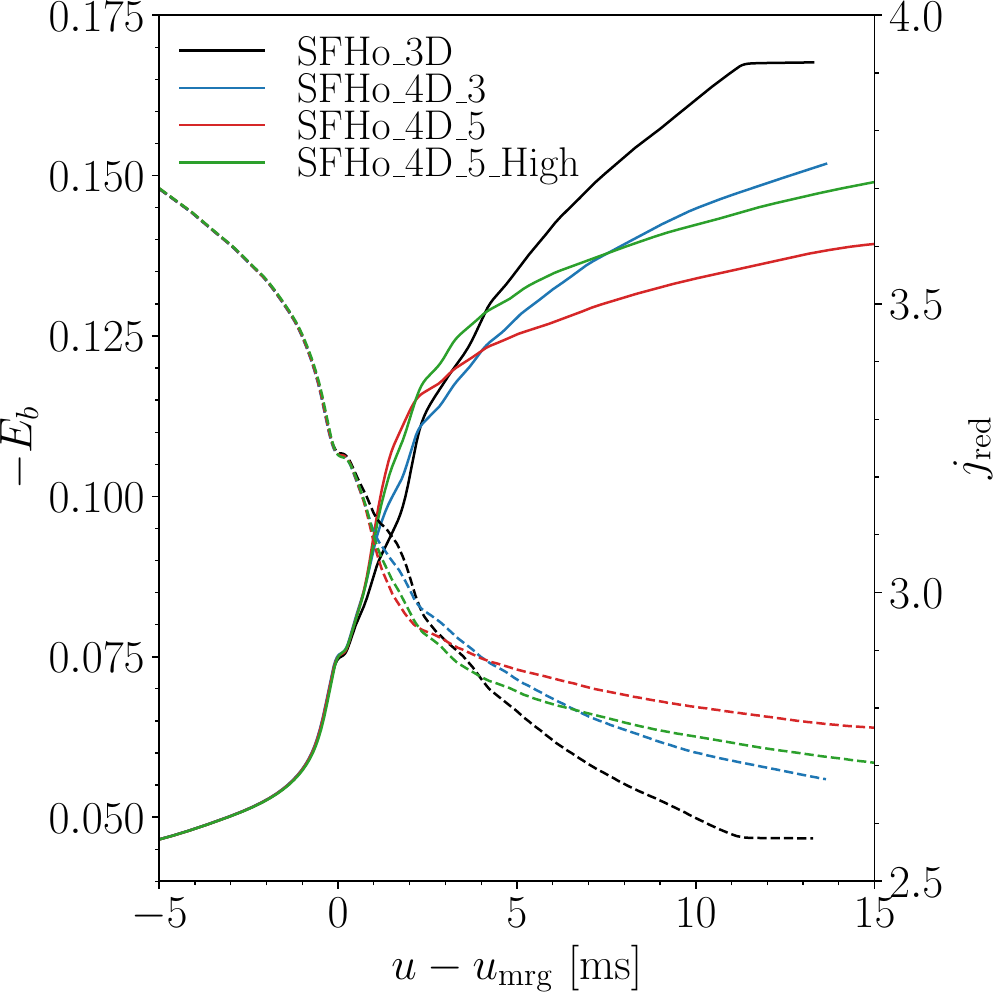}
    \end{subfigure}
    \caption{Evolution of the gravitational binding energy $-E_b$ (thick lines) and reduced angular momentum $j_{\rm red}$ (dashed lines) for our simulated setups. GW contributions to both quantities are extracted at a sphere of constant coordinate radius $r = 1000~{M_\odot}$. We employ the retarded time $u$ to account for the time-of-flight of the signal to the extraction sphere.}
    \label{fig:eb-j}
\end{figure}

As depicted in Fig.~\ref{fig:eb-j}, the time evolution of $-E_b$ (thick lines) and $j_{\rm red}$ (dashed lines) agree until the merger for all runs. Up to $5~{\rm ms}$ after the collision, the muonic runs emit more energy in the form of GWs, following the larger central density oscillations. Within this time span, the muonic systems are more compact, as expected from the softening of the EoS in the presence of muons, and SFHo$\_$4D$\_$5 is more compact than SFHo$\_$4D$\_$3, because some muonization occurs between the fusing cores (more details in Sec.~\ref{sec:muon-content}). Hence, the muonic setups undergo a more violent merger event. In this early stage, the evolution at high densities is dominated by GW emission, thus the energetics of neutrino emission is of little dynamical importance. At $5~{\rm ms}$ the remnant cores reach approximately the same central density, and we note the formation of a trend in $-E_b$, namely that SFHo$\_$3D is more strongly bound than SFHo$\_$4D$\_$3, that is more strongly bound than SFHo$\_$4D$\_$5. On the other hand, the ordering is reversed for $j_{\rm red}$ and for the retained mass within the remnants of each setup, suggesting that a stronger bounce efficiently distributes mass and angular momentum outwards. Accordingly, density oscillations are weakened, which in turn reduce energy and angular momentum losses via GWs. The remnant becomes less bound (higher $-E_b$), and a stronger rotational support (higher $j_{\rm red}$) prevents further contraction of the remnant. Indeed, the residual angular momentum of SFHo$\_$4D$\_$5 is enough to stabilize the remnant (at least up to our final simulation time), while for SFHo$\_$4D$\_$3 the smaller $j_{\rm red}$ only delays the collapse. At increased grid resolution, we note that the remnant is more strongly bound and rotational support is weaker for SFHo$\_$4D$\_$5$\_$High, leading to the longer-sustained oscillations and larger central rest-mass density when compared to SFHo$\_$4D$\_$5.

\begin{figure*}[ht!]
    \centering
    \includegraphics[width=\textwidth]{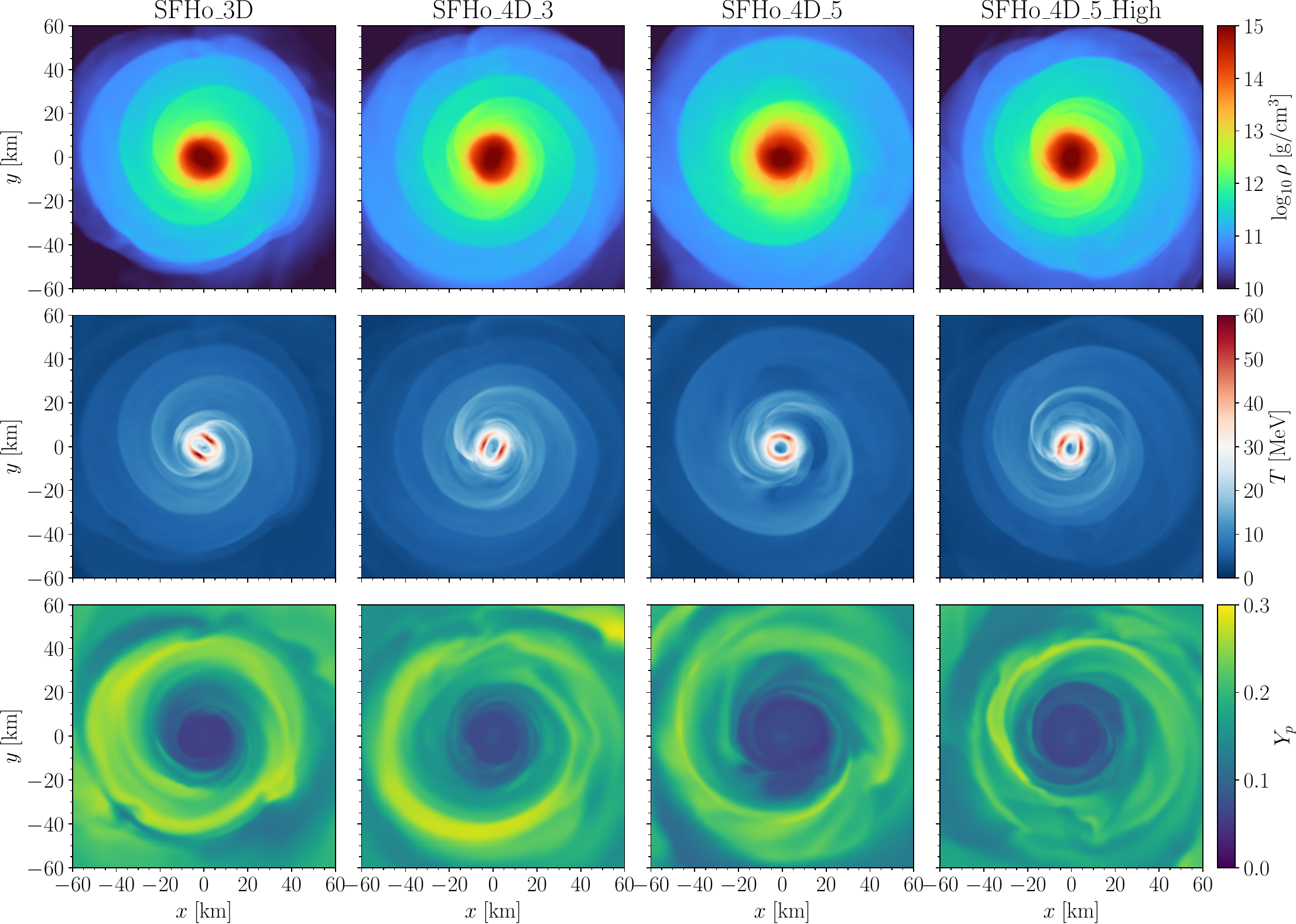}
    \caption{Matter properties in the $x-y$ plane for the SFHo runs at $t - t_{\rm mrg} = 10~{\rm ms}$. Columns are referred from left to right as first to fourth. \textit{First column:} SFHo$\_$3D. \textit{Second column:} SFHo$\_$4D$\_$3. \textit{Third column:} SFHo$\_$4D$\_$5. \textit{Fourth column:} SFHo$\_$4D$\_$5$\_$High. \textit{Upper row:} rest-mass density. \textit{Middle row:} temperature. \textit{Lower row:} proton fraction.}
    \label{fig:SFHo_xy_10ms}
\end{figure*}

Now we proceed with an analysis of the remnant and disk structure. In Fig.~\ref{fig:SFHo_xy_10ms} we depict the matter state on the $x-y$ plane $10~{\rm ms}$ after the merger. First we note that the SFHo$\_$3D setup develops the most compact disk (upper row, first column), with a hot core-disk interface, pronounced shocked-tidal arms (middle row, first column) and a highly protonized disk, with $Y_p \gtrsim 0.25$ up to $40~{\rm km}$ from the origin (lower row, first column). The SFHo$\_$4D$\_$3 run (second column), exhibits more pronounced tidally-shocked arms, although with overall smaller temperatures throughout the remnant core and disk, leading to a less protonized disk. It is worth noticing that the higher proton fraction in the densest portions of the muonic runs is reminiscent from the more protonized initial data (see Fig.~\ref{fig:Yp-beta-eq}).

The 5-species runs SFHo$\_$4D$\_$5 and SFHo$\_$4D$\_$5$\_$High (third and fourth columns) develop an extended $\rho \geq 10^{13}~{\rm g/cm^3}$ region, with a noticeable suppression of the formation of tidally shocked arms. Accordingly, the whole remnant and disk are cooler than for SFHo$\_$4D$\_$3 and SFHo$\_$3D and the proton fractions are smaller (lower row, third and fourth columns). In fact, weaker protonization is a direct consequence of the inclusion of muonic reactions, since the additional cooling by emission of muon-flavored neutrinos leads to a reduced electronization of matter via emission of electron-flavored neutrinos (a full presentation and discussion of neutrino luminosities are found in the next Section).

\begin{figure}
    \centering
    \includegraphics[width=0.94\columnwidth]{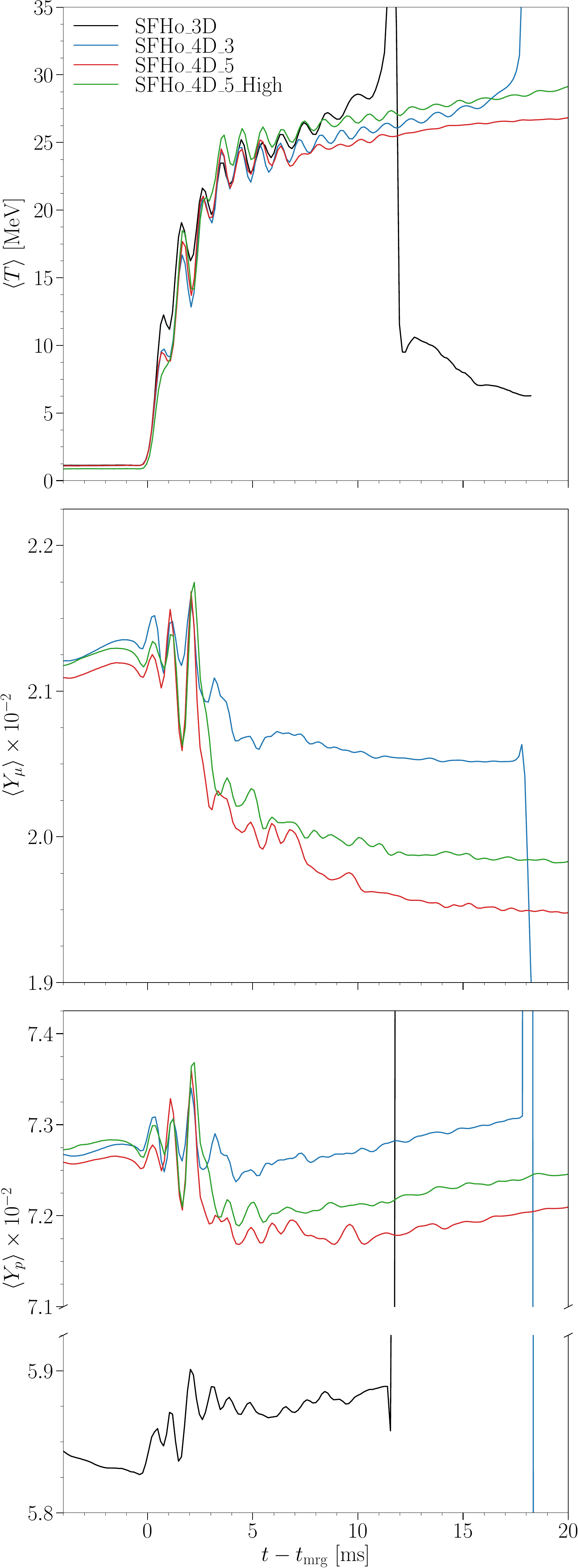}
    \caption{Evolution of the mass-averaged matter properties on the $x-y$ plane. Quantities are shown as moving time averages with window $0.5~{\rm ms}$. \textit{Upper panel:} average temperature, where we note that 3-species run reaches overall higher temperatures compared to the muonic runs at same grid resolution. \textit{Middle panel:} average muon fraction for the muonic runs. \textit{Bottom panel:} average proton fraction, where the higher values achieved during the inspiral by the muonic runs are reminiscent from the more protonized cores in the initial data.}
    \label{fig:TD-avg}
\end{figure}

Complementary to the snapshots presented in Fig.~\ref{fig:SFHo_xy_10ms}, we show in Fig.~\ref{fig:TD-avg} the time evolution of mass-averaged quantities of interest in the orbital plane.

We begin noting that the average temperature (upper panel) is overall higher for the SFHo$\_$3D run along the post-merger, which follows from a less potent neutrino cooling (see the luminosities presented in the next section). Even with the same set of weak reactions, SFHo$\_$4D$\_$3 is colder due to an increase of $\sim 10\%$ of the specific heat capacity at constant pressure under relevant conditions when muons are included. This increase is somewhat expected, given the enhanced energy cost of heating up a matter element containing one additional (massive) leptonic species. Furthermore, we verified that this increase is dominated by leptonic contributions, because changes in the baryonic contribution to the heat capacity due to the increased proton fraction in the presence of muons is very modest. In addition to that, in the SFHo$\_$4D$\_$5 run we observe that cooling provided by muonic weak reactions contributes to lowering the average temperatures and to the suppression of temperature oscillations at later times. Finally, the higher average temperature observed in the SFHo$\_$4D$\_$5$\_$High run is a resolution effect, related to a better capture of shocks during the post-merger.

In the following, we present a brief discussion about the evolution of the average muon fraction (middle panel), while in Sec.~\ref{sec:muon-content} we present in more details the relevant microphysics that determines the muonic content found in our simulations. In the SFHo$\_$4D$\_$3 run, we observe peaks (up to $2~{\rm ms}$ after the merger) with approximately same $\langle Y_\mu \rangle \sim 0.0215$, associated to the oscillations of the remnant. Since in this run there are no CC muonic reactions, the total number of muons is conserved, and the subsequent decrease is set by the advection of muons from the core towards the disk. When CC muonic reactions are taken into account, as seen for SFHo$\_$4D$\_$5($\_$High), the peaks become more pronounced, indicating episodes of alternating muonization and de-muonization. Then, at later times, the average muon fraction decreases substantially as a consequence of de-muonization of intermediate-low density material. It is worth pointing out that the smaller value for SFHo$\_$4D$\_$5 compared to SFHo$\_$4D$\_$3 before the merger results from de-muonization already at the inspiral stage, being the same effect at play during the late time de-muonization.

Finally, the average proton fraction (bottom panel) increases along the post-merger for SFHo$\_$3D, which is expected from the very well known electronization of matter under post-merger conditions. For the muonic runs, the average proton fraction is larger, as set by the initial data, having similarities with the evolution of the average muon fraction due to local charge neutrality, most notable in the first $2~{\rm ms}$ of the post-merger. At later times we observe a steady increase in response to the electronization of matter, with a slower growth for SFHo$\_$4D$\_$5($\_$High).

We end this Section by remarking that the remnant structures of the muonic runs, e.g. extended remnant and disks, are fully consistent with our analysis in terms of binding energy and angular momentum. Future work employing different EoSs will be necessary to understand whether the observed stabilization via rotational support is general.

\subsection{Neutrino Emission}\label{sec:nu-emiss}
\begin{figure*}[ht!]
    \centering
    \begin{subfigure}
    \centering
    \includegraphics[width=0.48\linewidth]{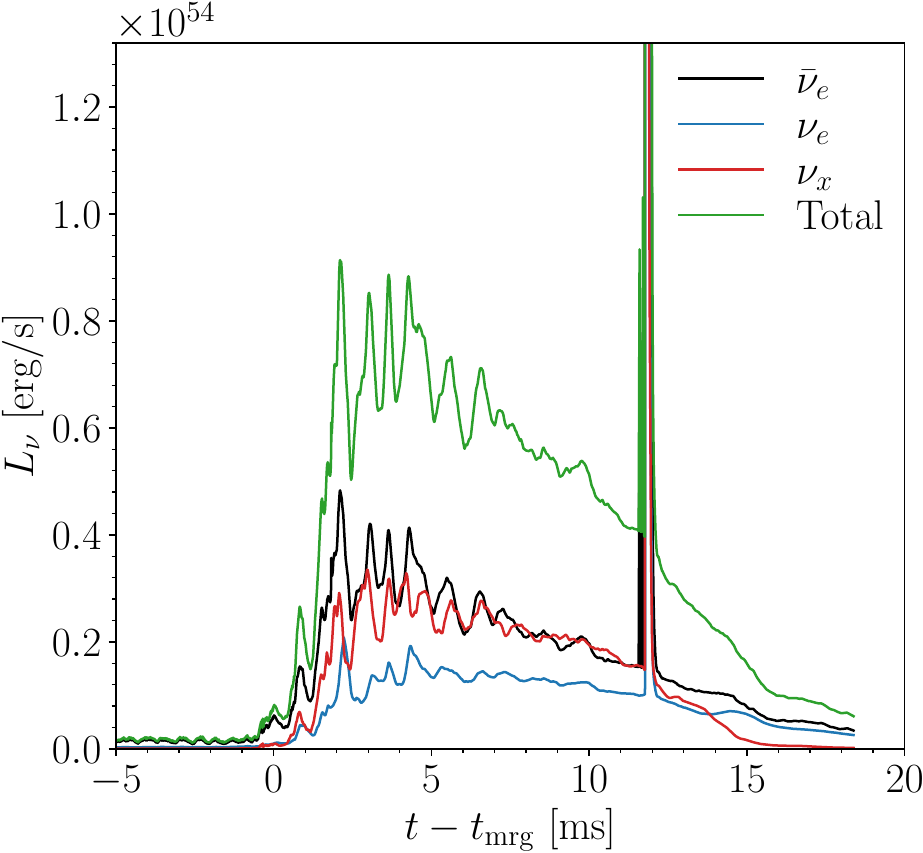}
    \end{subfigure}
    \begin{subfigure}
    \centering
    \includegraphics[width=0.48\linewidth]{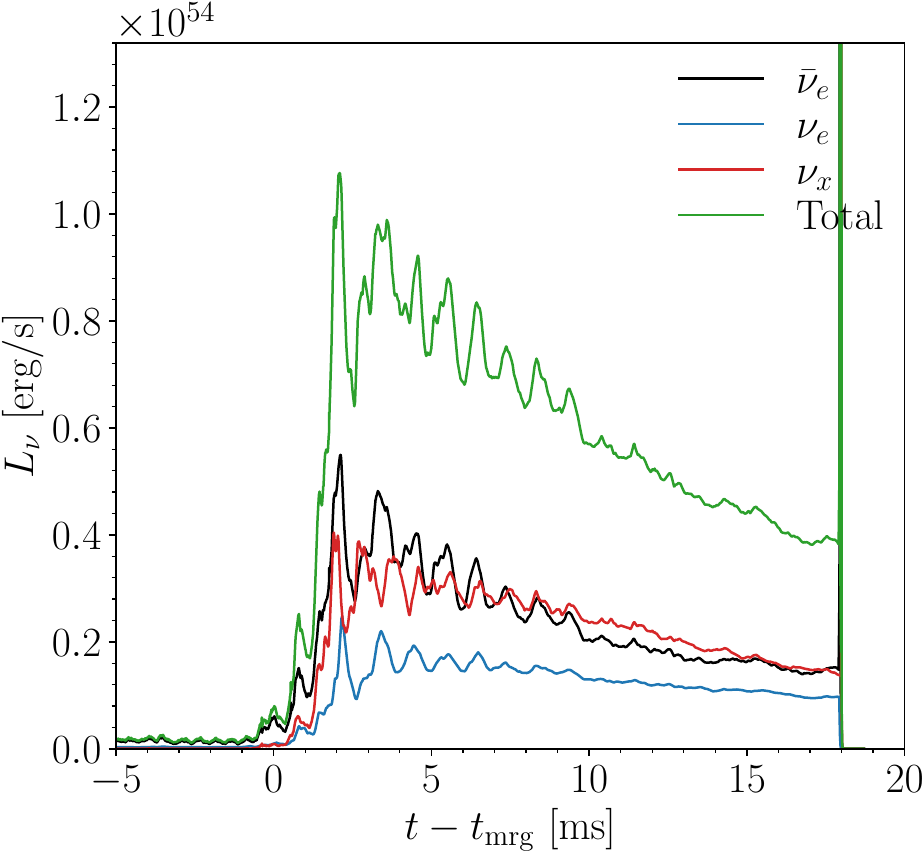}
    \end{subfigure}
    \begin{subfigure}
    \centering
    \includegraphics[width=0.48\linewidth]{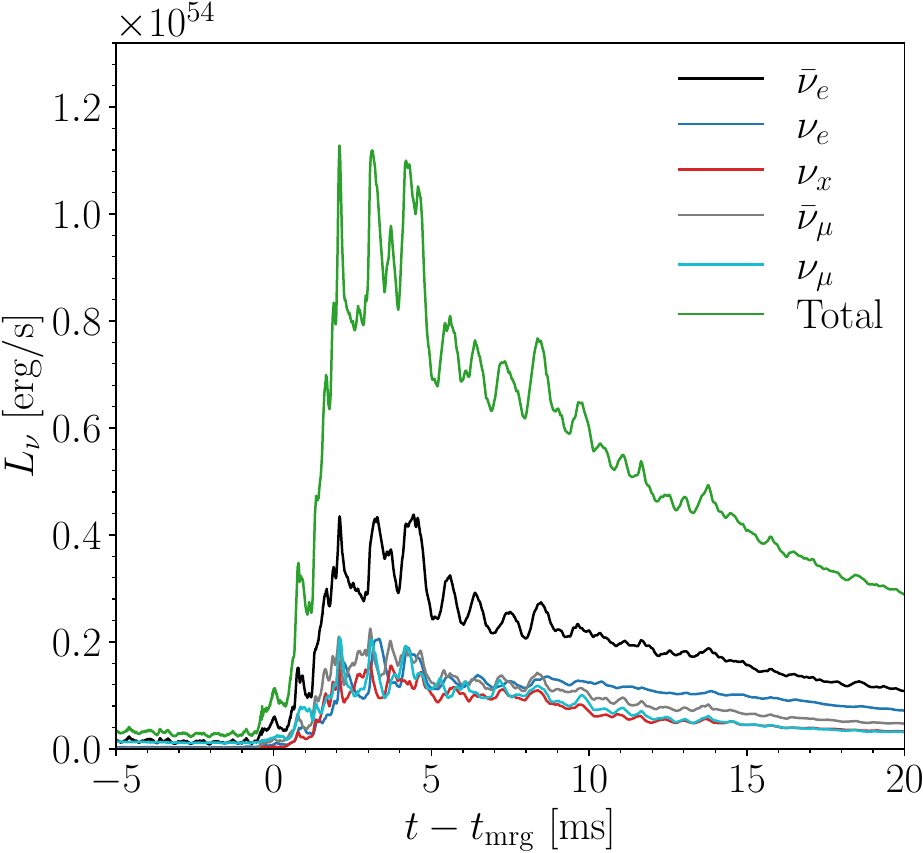}
    \end{subfigure}
        \begin{subfigure}
    \centering
    \includegraphics[width=0.48\linewidth]{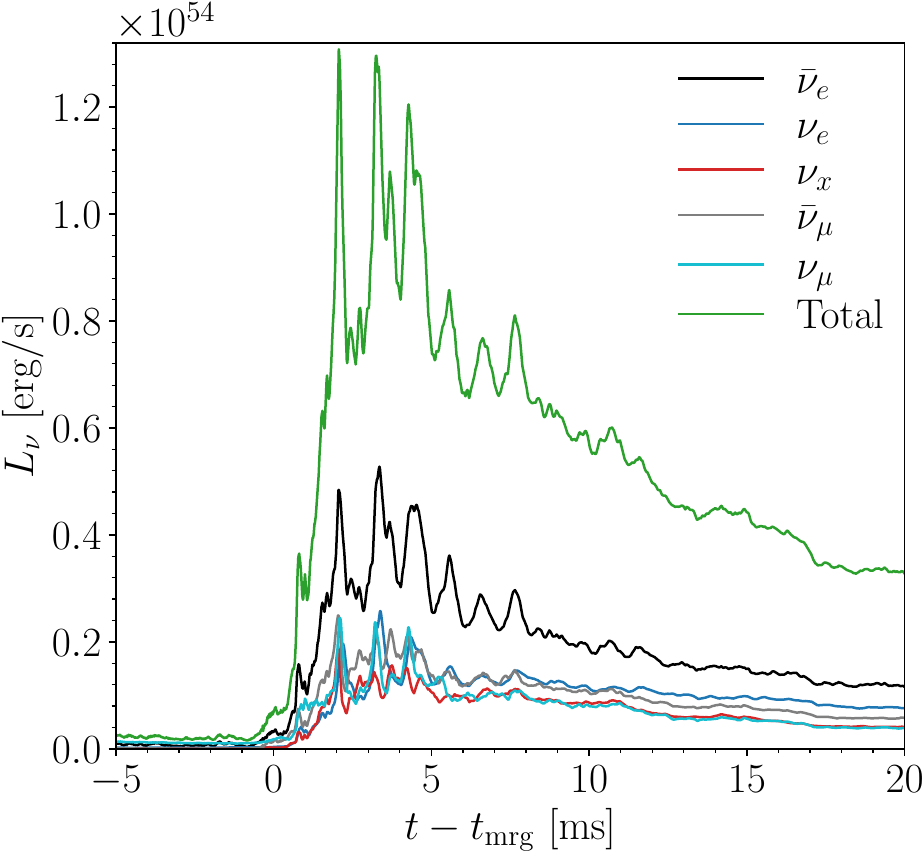}
    \end{subfigure}
    \caption{Source neutrino luminosity $L_\nu$ evolution. \textit{Upper left panel:} SFHo$\_$3D, where the peak around $12~{\rm ms}$ is related to the gravitational collapse. \textit{Upper right panel:} SFHo$\_$4D$\_$3, where we followed the simulation up to $1~{\rm ms}$ after the collapse for the purpose of comparing the remnant evolution with the remaining setups. \textit{Lower left panel:} SFHo$\_$4D$\_$5. \textit{Lower right panel:} SFHo$\_$4D$\_$5$\_$High. In all setups we note a burst of neutrinos shortly after the merger, prompted by the heating that follows the compression of matter elements. Before the merger, neutrinos are produced due to artificial heating produced by shocks in the interface between the stars and the atmosphere, although to a lesser degree with increased grid resolution.}
    \label{fig:nls_lum_SFHo}
\end{figure*}

In Fig.~\ref{fig:nls_lum_SFHo} we present the evolution of the neutrinos source luminosities. The pronounced peaks in the upper panels are associated to the gravitational collapse. Overall we note that the emissions peak around $\sim 2~{\rm ms}$ for all setups. The higher total luminosity peak of SFHo$\_$4D$\_$3 (upper right panel) when compared to the SFHo$\_$3D (upper left panel) at this instant follows from the higher $\rho_{\rm max}$ reached by the former (see Fig.~\ref{fig:rho_max}). In this case, the compression of matter increases the reaction rates across all neutrinos species.

Along the post-merger stage, $\bar\nu_e$ dominates up to $6~{\rm ms}$, followed by $\nu_x$ for the 3-species runs SFHo$\_$3D and SFHo$\_$4D$\_$3. After that, $L_{\nu_x}$ is otherwise comparable to $L_{\bar\nu_e}$ because of the high temperatures reached by the remnants and the strong temperature scaling of pair processes yielding $\nu_x$. Note that for the 5-species simulations, $L_{\nu_x}$ is smaller than in the 3-species counterparts, because in the later, $\nu_x$ represents the four species $\{\nu_\mu, \bar\nu_\mu,\nu_\tau, \bar\nu_\tau\}$, hence the effective rates contain statistical weight $4$, while in the former, $\nu_x$ represents $\{\nu_\tau,\bar\nu_\tau\}$ with statistical weight $2$ (see Sec.~\ref{sec:Methods}). Direct comparison shows that, $L_{\nu_\mu} + L_{\bar\nu_{\mu}} +L_{\nu_x} > L_{\nu_x}^{\rm 3-species}$, which suggests that the additional cooling provided by muonic $\beta$-processes in the 5-species scenario removes thermal energy from the fluid that would power the emission of heavy-lepton neutrinos via pair-processes. This same effect leads to the overall smaller luminosities observed across all neutrino species.
It is worth pointing out that during the formation of a black hole, we do not adopt any particular excision strategy. Instead, we let the rest-mass density evolve and linearly extrapolate thermodynamical quantities for densities above the maximum tabulated one. Such a procedure is somewhat arbitrary, but since this only happens within the apparent horizon (hence causally disconnected from the remaining of the grid), no significant effect is observed in the matter properties outside of the apparent horizon. However, since the optical depths computation depends on neighboring points, which might include points within the apparent horizon, unphysical optical depths may develop as a consequence of the linear extrapolation of opacities in those regimes. This is what we observe after the collapse for SFHo$\_$4D$\_$3: the maximum rest-mass density reaches between two and three times the maximum tabulated value, which produces unphysically high opacities and, consequently, very high optical depths within the apparent horizon. The optical depths at neighboring points then increase in response, leading to very small effective emission rates Eqs.~\eqref{eq:Qeff},~\eqref{eq:Reff} and source luminosities. Contrary, for the SFHo$\_$3D run, the maximum rest-mass density within the apparent horizon is $\sim 50 \%$ larger than the maximum tabulated value, thus the opacities do not reach as high of values and the optical depths are not contaminated by the region within the apparent horizon. Therefore, we are able to capture the fading luminosity emitted by the disk.
For the 5-species runs SFHo$\_$4D$\_$5 (lower left panel) and SFHo$\_$4D$\_$5$\_$High (lower right panel), we observe an overall larger total luminosity, which we associate to the additional cooling channels provided by the CC muonic reactions. The early post-merger neutrinos burst is such that the luminosities for $\nu_e$, $\nu_\mu$ and $\bar\nu_\mu$ are comparable up to $\sim 5~{\rm ms}$. Once the emissions stabilize (from around $10~{\rm ms}$ on), $L_{\nu_\mu} < L_{\bar\nu_\mu}$ because the hot and dense remnant is essentially optically thick and neutrinos mostly diffuse with average energies $\langle E_{\bar\nu_\mu} \rangle > \langle E_{\nu_\mu} \rangle$, as suggested by the average neutrino energies presented in Tab.~\ref{tab:avg-en}. We note that here the average neutrino energy is estimated as the ratio between energy and particle source luminosities, defined as in Ref.~\cite{Galeazzi:2013mia}, thus based on volume integrals over a grid level and without gravitational redshift corrections. Naturally, our reported average energies are higher by $\sim 50\%$ for all species when compared to the literature (e.g. Refs.~\cite{Foucart:2016rxm, Radice:2021jtw, Cusinato:2021zin, Schianchi:2023uky}), which is an expected feature for leakage schemes given that neutrino luminosity and energy estimates include neutrinos in the hot and dense remnant. On the other hand, more advanced treatments include absorption and neutrino properties are extracted far from the remnant, hence in the freely-streaming regime. Additionally, it is important to note that neutrino-electron scattering (NES) might play an important role for the thermalization of muon and tau-flavored neutrinos (see, e.g.,~\cite{Thompson:2000gv}). If included, one expects smaller $\langle E_{\nu_\mu} \rangle$, $\langle E_{\bar\nu_\mu} \rangle$ and $\langle E_{\nu_x} \rangle$ due to loss of neutrino energy in such inherently inelastic processes. Therefore, considering our simplified approach, our estimates should be regarded as semi-quantitative.
\begin{table}[ht!]
 \caption{Average neutrino energy per species, in MeV, $10~{\rm ms}$ after the merger for our simulated setups. The columns read simulation name, average electron neutrino energy, average anti-electron neutrino energy, average muon neutrino energy, average anti-muon neutrino energy and average heavy lepton-neutrino energy.}
    \centering
    \begin{tabular}{lccccc}
    \toprule
     Simulation    &  $\langle E_{\nu_e} \rangle$ & $\langle E_{\bar\nu_e} \rangle$ & $\langle E_{\nu_\mu} \rangle$ & $\langle E_{\bar\nu_\mu} \rangle$ & $\langle E_{\nu_x} \rangle$\\
       \hline \hline
                SFHo$\_$3D & $15.3$ & $22.2$ & $-$ & $-$ & $34.8$ \\
         SFHo$\_$4D$\_$3 & $15.8$ & $21.5$ & $-$ & $-$ & $33.7$ \\
         SFHo$\_$4D$\_$5 & $15.0$ & $21.5$ & $29.4$ & $44.1$ & $34.0$\\
         SFHo$\_$4D$\_$5$\_$High & $15.2$ & $22.0$ & $28.6$ & $40.3$ & $34.1$\\
         \hline \hline
    \end{tabular}
    \label{tab:avg-en}
\end{table}

Nevertheless, for all cases we recover the usual hierarchy $\langle E_{\nu_x} \rangle > \langle E_{\bar\nu_e} \rangle > \langle E_{\nu_e} \rangle$. Irrespective to the presence of muons and muonic weak reactions, $\langle E_{\nu_e} \rangle$, $\langle E_{\bar\nu_e} \rangle$ and $\langle E_{\nu_x} \rangle$ agree within $\sim 5\%$. Interestingly, for the 5-species runs we note that $\sim 10~{\rm ms}$ after the merger, the neutrino-spheres for $\bar\nu_\mu$, $\nu_x$ and $\nu_\mu$ are, respectively, located at increasing radii from the remnant center, although somewhat close, and are found deeper within the remnant, hence at higher matter temperatures, than the $\bar\nu_e$ and $\nu_e$ neutrino-spheres. Thus, our results are in qualitative agreement with the conclusion of Ref.~\cite{Cusinato:2021zin} that the observed energy hierarchy is related to the higher temperature of the medium from which $\nu_x$, $\nu_\mu$ and $\bar\nu_\mu$ decouple, with a caveat that there the authors employ an M0 scheme for the transport of energy and particle number, hence approximately accounting for neutrino absorption. Besides, since the emission rates for pair processes are the same for $\nu_\mu$ and $\bar\nu_\mu$, their difference in average energies comes from the CC reactions. In fact, by enforcing the lower bound on the energy of neutrinos that may be emitted via CC processes, we have $E_{\bar\nu_\mu,\min} = m_\mu c^2 + Q \approx 107 - 170~{\rm MeV}$, while $E_{\nu_\mu,\min} = m_\mu c^2 - Q \approx 45 - 104~{\rm MeV}$.
Our results suggest the following neutrino energy hierarchy: $\langle E_{\bar\nu_\mu} \rangle > \langle E_{\nu_x} \rangle > \langle E_{\nu_\mu} \rangle > \langle E_{\bar\nu_e} \rangle > \langle E_{\nu_e} \rangle$, although further simulations employing more EoSs and an improved neutrino treatment are desirable to draw firmer conclusions.

\subsection{Muon Content}\label{sec:muon-content}
\begin{figure*}[ht!]
    \centering
    \includegraphics[width=\textwidth]{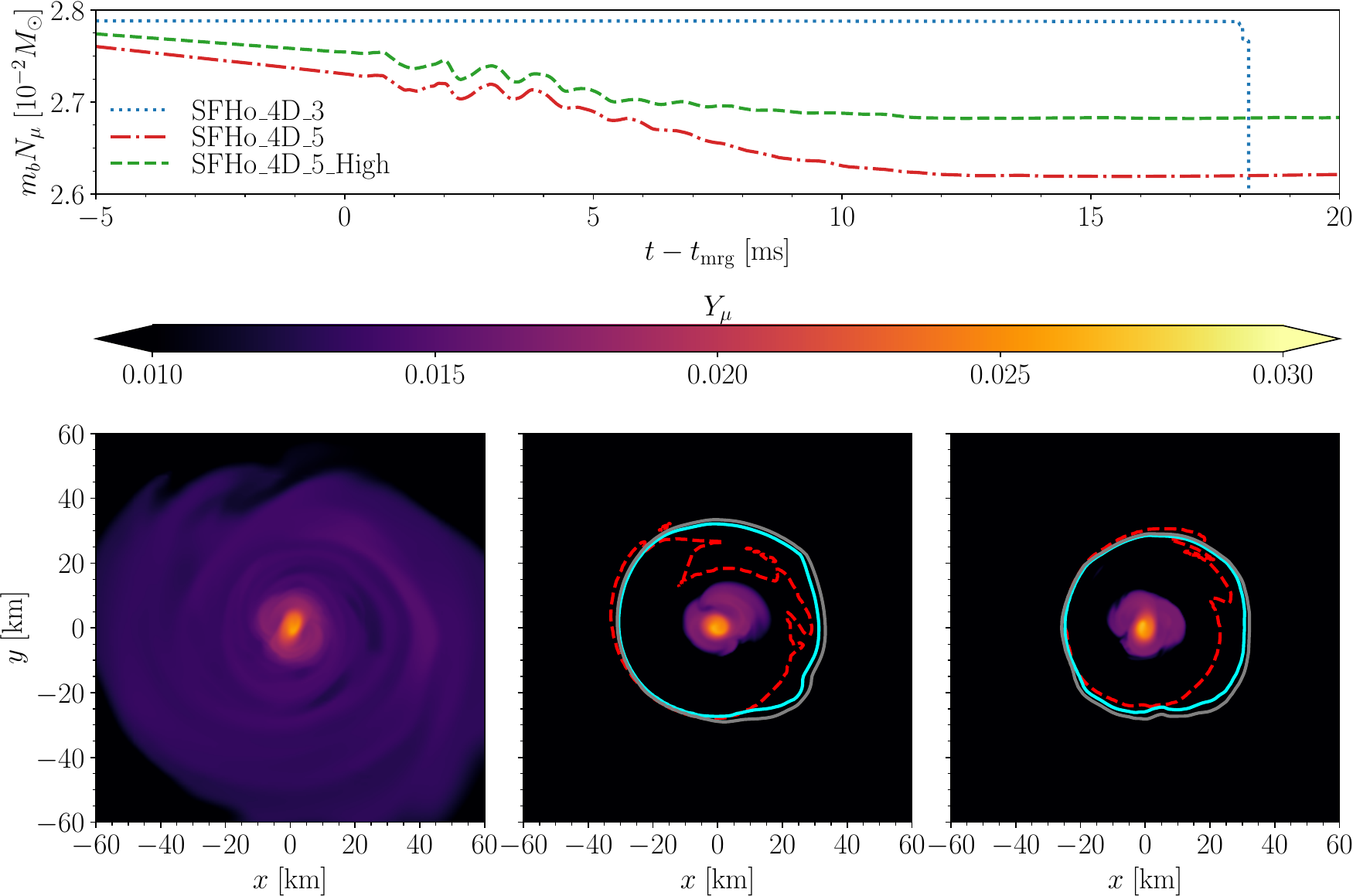}
    \caption{Evolution of the muon fraction for the muonic SFHo runs. \textit{Upper panel:} Evolution of the conserved number of muons in the grid level $L=1$. The 3-species run SFHo$\_$4D$\_$3 conserves $m_b N_\mu$ along the evolution, while the 5-species exhibits de-muonization of matter. Here we observe that such effect takes place before the merger due to the expansion of low density, cold material into the atmosphere, but to lesser extent with increased resolution.  \textit{Lower panels:} $Y_\mu$ on the $x-y$ plane $10~{\rm ms}$ after the merger. \textit{Lower left panel:} SFHo$\_$4D$\_$3 run, where the region with $Y_\mu > 0.01$ extends up to $60~{\rm km}$ from the origin. \textit{Lower middle panel:} SFHo$\_$4D$\_$5 run, where the disk is found de-muonized, while $Y_\mu \gtrsim 0.015$ is present in regions where $\rho \geq 10^{14}~{\rm g~cm^{-3}}$. \textit{Lower right panel:} SFHo$\_$4D$\_$5$\_$High run. The red dashed line marks $Y_\mu = 10^{-6}$ and the thick contour lines mark the neutrino-spheres of $\nu_\mu$ (cyan) and $\bar\nu_\mu$ (gray), where $\tau_{\nu_\mu, 0} = \tau_{\bar\nu_\mu, 0} = 1$.}
    \label{fig:SFHo_DYm_int}
\end{figure*}

In order to describe the evolution of the muonic content we introduce the conserved muon number, defined as
\begin{equation}
    m_b N_{\mu} = \int_{\mathcal{V}}  DY_\mu \sqrt{\gamma} d^3x,
\end{equation}
where $D = W\rho$ is the rest-mass density in the Eulerian frame, $W$ is the usual Lorentz factor and the integration volume $\mathcal{V}$ is a grid level. In the absence of muonic weak reactions, the balance-law Eq.~\eqref{eq:m-cons} implies the approximate constancy of $N_\mu$ along the evolution, which is verified for SFHo$\_$4D$\_$3 up to the collapse in the upper panel of Fig.~\ref{fig:SFHo_DYm_int}. In contrast, when CC muonic reactions are included (SFHo$\_$4D$\_$5($\_$High) runs), we observe that the conserved muon number decreases by as much as $8 \%$ ($4 \%$) with respect to the initial condition within our simulation time span. In fact, we observe an early de-muonization, prompted by the expansion of cold, low density material into the artificial atmosphere (more details in the following). However, such an effect is diminished both before and after the merger with increased grid resolution, because the hydrodynamics is better resolved and less matter expands into the atmosphere.

The de-muonization can be visualized in the lower panels of Fig~\ref{fig:SFHo_DYm_int}, where we show $Y_\mu$ in the $x-y$ plane $10~{\rm ms}$ after the merger. 
For the SFHo$\_$4D$\_$3 run (lower left panel), we note that a substantial $Y_\mu > 0.01$ is distributed throughout the disk, which is a hydrodynamical effect attributed to the sourceless advection of muons from the remnant core. Contrary, for the SFHo$\_$4D$\_$5 (lower middle panel) and SFHo$\_$4D$\_$5$\_$High (lower right panel) muons are found only within the remnant, where $\rho \geq 10^{14}~{\rm g~cm^{-3}}$ (see third and fourth, left panels of Fig.~\ref{fig:SFHo_xy_10ms}), rapidly decreasing outside of this region, as indicated by the $Y_\mu = 10^{-6}$ red dashed contour line.

\begin{figure}[ht!]
    \centering
    \includegraphics[width=0.48\textwidth]{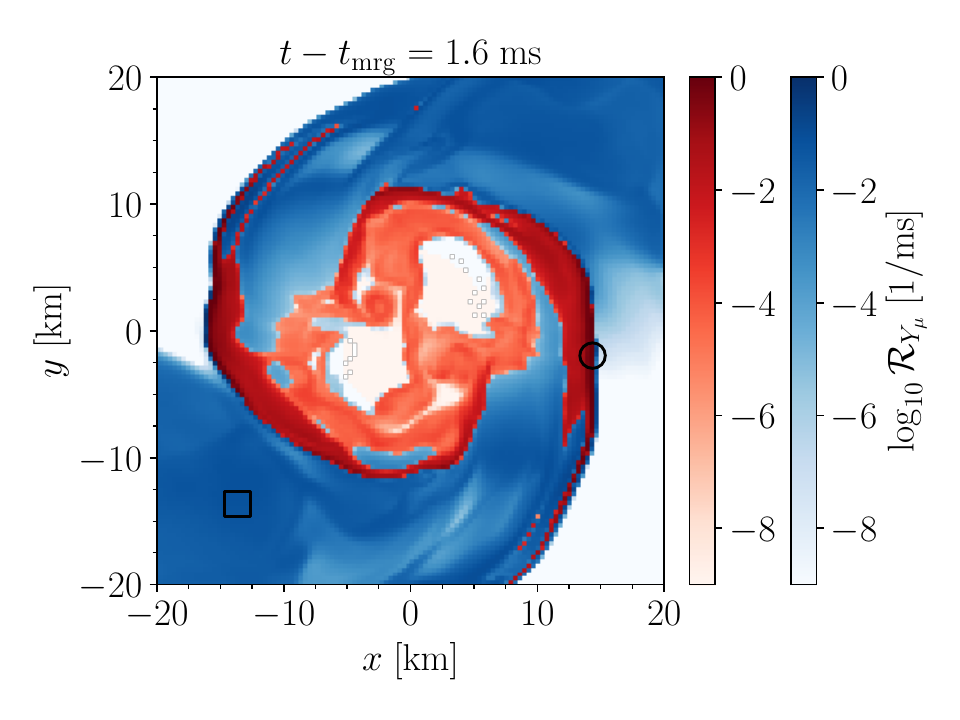}
    \caption{Effective muon production rate due to emission of muon-flavored neutrinos for the SFHo$\_$4D$\_$5 run at $t - t_{\rm mrg} = 1.6~{\rm ms}$, roughly coinciding with the increase in $m_bN_\mu$ depicted in the upper panel of Fig.~\ref{fig:SFHo_DYm_int}. The red colorbar represents regions where $\mathcal{R}_{Y_\mu} >0$ (muonization), while the blue colorbar represents regions where $\mathcal{R}_{Y_\mu} <0$ (de-muonization). For reference, we mark with a black circle a point with high muonization rate, and a square marks high de-muonization rate.}
    \label{fig:effective-rates}
\end{figure}

For a better understanding of the conditions in which (de)muonization operates, we present in Fig.~\ref{fig:effective-rates} the effective muon production rate $\mathcal{R}_{Y_\mu}$
\begin{equation}
    \mathcal{R}_{Y_\mu} = \frac{\alpha m_b}{W\rho}(R^{\rm eff}_{\bar\nu_\mu} - R^{\rm eff}_{\nu_\mu}),
\end{equation}
which is simply the r.h.s of Eq.~\eqref{eq:m-cons} combined with the baryon number conservation Eq.~\eqref{eq:bar-cons}, and represents the time rate of change of $Y_\mu$ by the emission of muon-flavored neutrinos. Hence, $\mathcal{R}_{Y_\mu} >0$ (red colorbar) corresponds to muonization, $\mathcal{R}_{Y_\mu} < 0$ (blue colorbar) corresponds to de-muonization and white regions correspond to negligible production rates. It is clear that muonization is more intense in the hot, medium-density, shocked regions, represented by the point marked with a circle, where $\mathcal{R}_{Y_\mu} > 10^{-2}/{\rm ms}$ and $\rho = 2.7\times10^{12}~{\rm g/cm^3}$, $T = 21.4~{\rm MeV}$, $Y_p = 0.14$ and $Y_\mu = 0.024$. On the other hand, in the innermost portions of the remnant, where densities are higher, muonization proceeds much more slowly ($10^{-3} - 10^{-6}/{\rm ms}$), allowing a modest muon buildup in the shear heated region between the fusing cores. Furthermore, the slower observed muonization confirms the findings of Ref.~\cite{Loffredo:2022prq}, that the bulk of muons found in the interior of the remnant at later times are reminiscent from the initial cold cores.

In contrast, de-muonization is observed predominantly in cold, medium-to-low density matter streams. The square marker in Fig.~\ref{fig:effective-rates} represents a point where $|\mathcal{R}_{Y_\mu}| > 10^{-2}/{\rm ms}$, $\rho = 4.8\times10^{11}~{\rm g/cm^3}$, $T = 6.8~{\rm MeV}$, $Y_p = 0.16$ and $Y_\mu = 1.63\times10^{-6}$. Therefore, during the remnant's evolution, it is expected that matter elements in the outermost regions will decompress and cool, eventually reaching thermodynamic conditions where de-muonization operates fast.

\begin{figure}[ht!]
    \centering
    \includegraphics[width=0.4\textwidth]{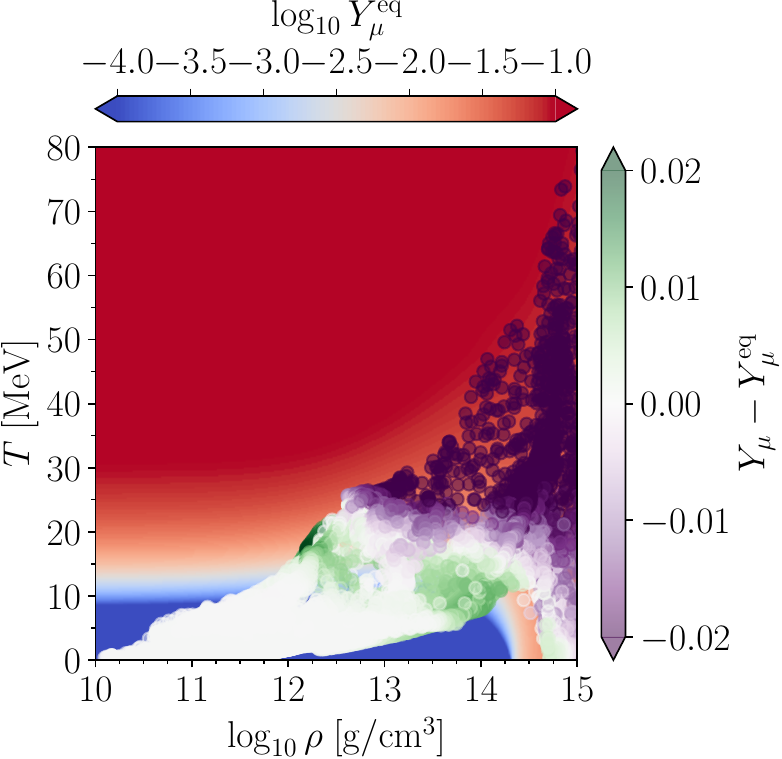}
    \caption{Neutrinoless $\beta$-equilibrium values $Y_\mu^{\rm eq}$ of the muon fraction with respect to $(\rho,T)$ (blue-red colorbar) and deviation $Y_\mu - Y_\mu^{\rm eq}$ of simulated $Y_\mu$ with respect to equilibrium (purple-green colorbar). In the diagram we see that at temperratures $T \lesssim 15~{\rm MeV}$ and low densities simulated values are found in equilibrium, marked by white points. Out-of-equilibrium matter elements are expected to equilibrate either by de-muonization (green points), found at intermediate densities and temperatures, or muonization (purple points), located at $T\gtrsim 15~{\rm MeV}$.}
    \label{fig:beta-eq}
\end{figure}

In order to have a qualitative understanding of the thermodynamics of de-muonization, we present in Fig.~\ref{fig:beta-eq} the muon fraction in neutrinoless $\beta$-equilibrium $Y_\mu^{\rm eq}$ (blue to red colorbar), obtained by solving Eqs.~\eqref{eq:ch-neut},~\eqref{eq:beta-eq} over the whole EoS subspace $(\rho, T)$. We stress out that the neutrinoless hypothesis is not particularly accurate at high densities, since neutrinos are expected to be in the trapped regime~\cite{Perego:2019adq}, but it is a suitable approximation to model the equilibration in a low-density environment, where neutrinos should be freely streaming.

The diagram encodes most of the relevant information regarding the $\beta$-equilibration process for muons, which we highlight in the following. First, at low temperatures, muons are present only at sufficiently high densities, e.g., where $\mu_e \geq m_\mu c^2$. At densities smaller than $\approx 1.8\times10^{14}~{\rm g/cm^3}$ and temperatures smaller than $\approx 10~{\rm MeV}$, the equilibrium fractions are negligible, but at higher temperatures $Y_\mu^{\rm eq}$ rapidly increases as the muon/antimuon gas becomes non-degenerate. In the purple-green colorbar we present the deviation of the muon fraction $Y_\mu$ in the $x-y$ plane at $t - t_{\rm mrg} = 1.6~{\rm ms}$ with respect to the $\beta$-equilibrium value at the densities and temperatures encountered in our simulation. There it becomes clear that cold ($T \leq 10~{\rm MeV}$) matter elements with $\rho \leq 10^{12}~{\rm g/cm^3}$, corresponding to the typical thermodynamical conditions where higher de-muonization rates are observed in Fig.~\ref{fig:effective-rates}, are not only effectively de-muonized, but also that such de-muonization is a consequence of $\beta$-equilibration. This phenomenology plausibly explains the early de-muonization during the inspiral.

Green points represent matter elements with muon fraction in excess, hence expected to de-muonize, and are predominantly distributed in a region $10^{12}~{\rm g/cm^3} \leq \rho \leq 2\times10^{14}~{\rm g/cm^3}$ and temperatures smaller than $20~{\rm MeV}$.

Purple points represent matter elements expected to muonize, mainly found at temperatures higher than $20~{\rm MeV}$, and to a lesser extent at densities $\approx 10^{15}~{\rm g/cm^3}$. Those points are mapped to positive production rates (see Fig.~\ref{fig:effective-rates}), but since most of the muonization occurs very slowly, at $\mathcal{R}_{Y_\mu} < 10^{-4}/{\rm ms}$ (except for the shocked arms), the faster decompression, cooling and subsequent de-muonization dominate the muon fraction evolution up to the stationary state depicted in the upper panel of Fig.~\ref{fig:SFHo_DYm_int}.

Here we make some remarks about our findings. First, the muon fraction distributions along the $x-y$ plane at $t - t_{\rm mrg} = 5~{\rm ms}$ is similar to Fig.~3 of Ref.~\cite{Loffredo:2022prq}, i.e., around the final instants of the neutrino bursts, we found $Y_\mu = 10^{-4} - 10^{-2}$ within $25~{\rm km}$ of the recently formed remnant for the 5-species runs, which is due to the early redistribution of $Y_\mu$ from the merging cores. Thus, the de-muonization reported by us operates on longer timescales, mostly affecting the outermost disk regions. 

Next, it is worth pointing out that, contrary to the CCSNe simulations of Refs.~\cite{Bollig:2017lki, Fischer:2020vie}, we do not observe the muonization of high-density matter, which is explainable by the following: first, their simulations start without muons within the matter. Then a substantial $Y_\mu$ builds up only shortly before the core bounce and after it, most importantly through a two-stages process comprised of production of high-energy muon-(anti)neutrinos via pair processes that may then participate in muonic absorption reactions. Such a mechanism cannot be properly modeled by a neutrinos leakage scheme, because absorption is not realistically captured by the treatment.
Second, the protonization observed in NLS simulations is based on an excess emission of $\bar\nu_e$ with respect to $\nu_e$, which occurs in the intermediate region between spatially separated $\bar\nu_e$ and $\nu_e$ neutrino-spheres such that $\tau_{\bar\nu_e, 0} = 1$ is located closer to the remnant than $\tau_{\nu_e, 0} = 1$. This is not the case for $\nu_\mu$ and $\bar\nu_\mu$ neutrino-spheres. Instead, what we observe is that the $\bar\nu_\mu$ neutrino-sphere is slightly wider than the $\nu_\mu$ neutrino-sphere, as depicted in the lower panels of Fig.~\ref{fig:SFHo_DYm_int}. This happens because the spectrally-averaged opacities employed in this work are heavily dominated by scattering processes in the muon(anti)-neutrino neutrino-spheres. Besides, there the neutrino degeneracies are relatively small $\eta_{\nu_\mu} = -\eta_{\bar\nu_\mu} \approx -2$, such that the number-averaged opacity is typically a few tens of percent larger for $\bar\nu_\mu$ than for $\nu_\mu$. On the other hand, the maximum values of the optical depths found during the post-merger are around one order of magnitude smaller for $\bar\nu_\mu$ than for $\nu_\mu$, which is expected given that the energy-dependent CC opacities are one to two orders of magnitude smaller for $\bar\nu_\mu$ than for $\nu_\mu$ under remnant and disk conditions~\cite{Fischer:2020vie}. 
Compared to late stages of the post-bounce reported by Ref.~\cite{Fischer:2020vie}, we found $Y_\mu$ in excess of about one order of magnitude. This is because most of the $Y_\mu$ found in the remnant cores come from the initially cold, neutrinoless, $\beta$-equilibrated NSs, while in the former, matter is still hot ($T \approx 15~{\rm MeV}$) and far from $\beta$-equilibrium, i.e., there $\mu_n - \mu_p < \mu_\mu < \mu_e$. Finally, our $Y_\mu$ values are comparable to Ref.~\cite{Bollig:2017lki} under similar thermodynamic conditions, which also agrees with the reported in Refs.~\cite{Loffredo:2022prq, Pajkos:2025oyf}.

\section{Ejecta Analysis}
\label{sec:Ej}
In this Section we present an analysis of the ejecta properties for our simulations. The relevant quantities are summarized in Table~\ref{tab:ej_prop}. We note that the geodesic criterion~\cite{Hotokezaka:2013b} was adopted and the reported averages were extracted at a fixed sphere of coordinate radius $r = 300~M_\odot$ by the procedure outlined in Ref.~\cite{Schianchi:2023uky}.

\begin{table}[ht!]
 \caption{Ejecta properties for our simulated setups. Columns show the simulation, ejecta mass extracted at $r = 200~M_\odot$, ejecta mass extracted at $r = 300~M_\odot$, average proton fraction, average entropy per baryon and average velocity measured by an observer at infinity. All average quantities were extracted at $r = 300~M_\odot$. For the muonic runs, $\langle Y_\mu \rangle \lesssim 10^{-3}$, thus $\langle Y_p \rangle \approx \langle Y_e \rangle$.}
    \centering
    \begin{tabular}{lcccccc}
    \toprule
     Simulation    &  $M_{\rm ej}^{r = 200 M_\odot}$ & $M_{\rm ej}^{r = 300 M_\odot}$ & $\langle Y_p \rangle$ & $\langle s \rangle$ & $v_{\infty}$\\
         &  $[10^{-3} M_\odot]$ & $[10^{-3} M_\odot]$ & & [$k_B$] & [$c$] \\
       \hline \hline
                SFHo$\_$3D & $2.8$ & $2.2$ & $0.17$ & $11.3$ & $0.19$ \\
         SFHo$\_$4D$\_$3 & $1.9$ & $1.2$ & $0.20$ & $12.0$ & $0.17$\\
         SFHo$\_$4D$\_$5 & $1.5$ & $1.0$ & $0.16$ & $11.4$ & $0.16$ \\
         SFHo$\_$4D$\_$5$\_$High & $1.6$ & $1.0$ & $0.19$ & $13.6$ & $0.14$\\
         \hline \hline         
    \end{tabular}
    \label{tab:ej_prop}
\end{table}

We begin comparing our SFHo$\_$3D results with works that employ similar physical setups and neutrinos treatment. Our ejecta masses extracted at $r = 200~M_\odot$ are, respectively, $\approx 30 \% (\approx 23 \%)$ smaller than those in Ref.~\cite{Bovard:2017mvn} (Ref.~\cite{Lehner:2016lxy}). When comparing ejecta masses extracted at $r = 300~M_\odot$ with those of Ref.~\cite{Radice:2018pdn}, we have $\approx 37 \%$ less. Such differences may be partly explained by differences in the hydrodynamics implementations, e.g., Refs.~\cite{Radice:2018pdn, Bovard:2017mvn} employ a positivity-preserving limiter with the MP5 reconstruction and a different prescription for the atmosphere~\cite{Radice:2013xpa}. On the other hand, our approach for the computation of opacities and emissivities should also contribute, given that ejecta properties are importantly impacted by the treatment of neutrinos.

Regarding average properties, good agreement is found for $\langle Y_p \rangle$, although our $\langle s \rangle$ is smaller by $\sim 24~\%$ and $v_\infty$ is smaller by $\sim 27~\%$ compared to Ref.~\cite{Radice:2018pdn}.

Comparing our SFHo runs, the SFHo$\_$4D$\_$3 setup has a more protonized $\langle Y_p \rangle = 0.20 $ and slightly more entropic $\langle s \rangle = 12.0$ ejecta than SFHo$\_$3D, although less massive by a factor $1.5 - 1.8$. The smaller amount of ejecta is consistent with the larger total luminosity of the former compared to the later (see Fig.~\ref{fig:nls_lum_SFHo}), by means of which energetic matter elements become gravitationally bound due to neutrinos emission. The same rationale applies to the SFHo$\_$4D$\_$5 and SFHo$\_$4D$\_$5$\_$High setups, which exhibit even higher total luminosities because of the additional muonic charged-current cooling channels. It should be noted, however, that our interpretation does not rule out the possibility that pressure changes due to the presence of muons may also affect the ejecta mass, as pointed out in Ref.~\cite{Loffredo:2022prq}.

At same grid resolution, SFHo$\_$3D and SFHo$\_$4D$\_$5 have very similar $\langle Y_p \rangle$ and $\langle s \rangle$, with differences mainly in $v_\infty$ by a factor of $\sim 1.2$ and in ejecta masses by a factor of $1.9 - 2.2$. With increasing resolution we note that SFHo$\_$4D$\_$5 and SFHo$\_$4D$\_$5$\_$High vary by less than $10 \%$ in the ejecta masses and in less than $20 \%$ in $\langle Y_p \rangle$, $\langle s \rangle$ and $v_\infty$, allowing us to estimate numerical uncertainties of at least $20 \%$.

\begin{figure*}[ht!]
    \centering
    \includegraphics[width=\textwidth]{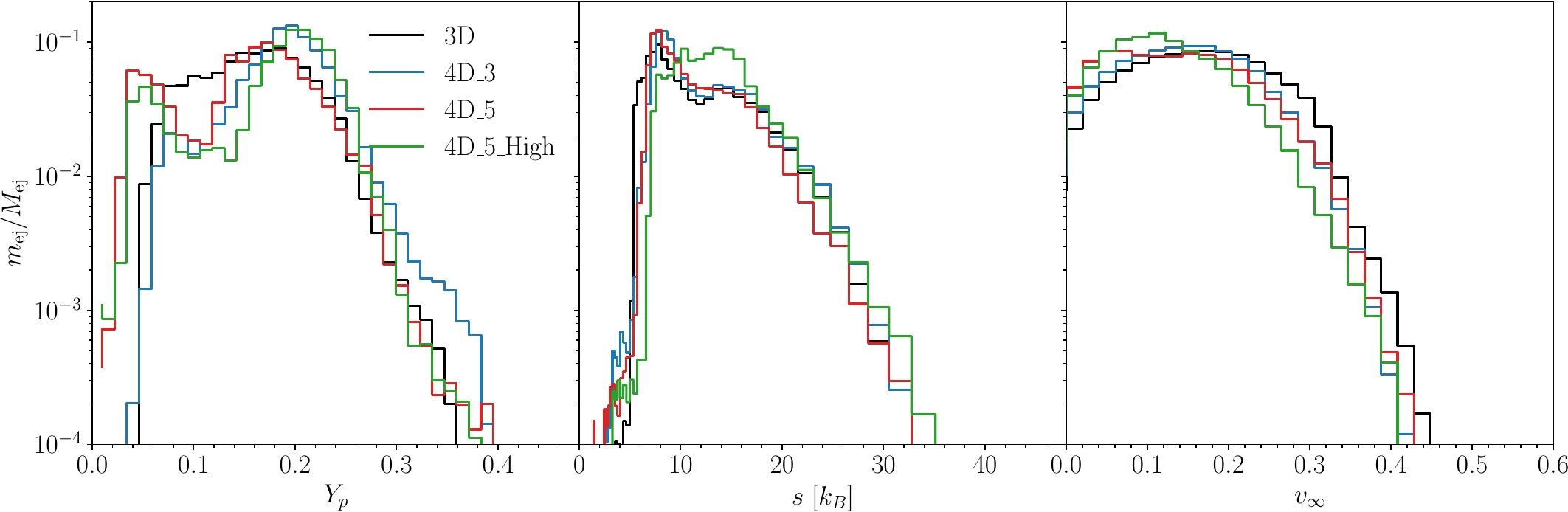}
    \caption{Distribution of ejecta mass detected at $r = 300~M_\odot$ with respect to the proton fraction $Y_p$ (left panel), entropy per baryon (middle panel) and asymptotic velocity (right panel).}
    \label{fig:yp-hist}
\end{figure*}

In Fig.~\ref{fig:yp-hist}, we present the distributions of ejected mass fractions with respect to the proton fraction $Y_p$ (left panel), entropy per baryon $s$ (middle panel) and asymptotic velocity $v_\infty$ (right panel) for our simulations, extracted at $r = 300~M_\odot$. Since at this position the muon fraction $Y_\mu$ for the muonic runs are much smaller than $Y_p$, the distributions are identical with respect to $Y_e$ and our results may be compared to others from the literature. Comparing our SFHo$\_$3D result (left panel) to similar runs of Refs.~\cite{Bovard:2017mvn, Radice:2018pdn}, our $Y_p$ distribution is flatter from $Y_p = 0.06$ up to the peak at $Y_p = 0.20$, followed by a similar fall-off for $Y_p \geq 0.3$. For SFHo$\_$4D$\_$3 the distribution is more clearly peaked at $Y_p = 0.20$, with considerably smaller fraction of neutron-rich material and a tail of neutron-poor material $Y_p \leq 0.30$. We verified that, in both simulations, the (neutron-rich) tidal component of the ejecta is rapidly reached by the (neutron-poor) shock-driven component of the ejecta, such that the material is reprocessed and the distribution is shifted towards higher $Y_p$ values.
In the case of SFHo$\_$4D$\_$5 and SFHo$\_$4D$\_$5$\_$High runs, the additional energy and momentum losses associated to the muonic reactions yield higher fractions of small velocity material (see right panel of Fig.~\ref{fig:yp-hist}). Hence, the reprocessing mechanism is inhibited and we note a pronounced neutron-rich secondary peak along with the expected dominant neutron-poor peak. It should be noted that the higher peak around $Y_p = 0.2$ and prolonged tail found for the muonic runs follow from the longer simulation times until collapse. Hence, matter with higher proton fraction have sufficient time to travel to the extraction sphere.
The entropy per baryon distribution (middle panel) is very similar across all setups at same grid resolution peaking at $s \sim 8~k_B$ and followed by a rapid decay such that a negligible fraction of the ejecta is found with $s \geq 32~k_B$. For the SFHo$\_$4D$\_$5$\_$High setup, the peak is shifted towards $s \sim 14~k_B$ and a tail is found up to $s \sim 35~k_B$. The asymptotic velocity (right panel) follows a similar pattern for all simulations with a trend that increased amounts of small velocity ejecta are found in runs with higher total neutrinos luminosities. Besides, we did not find in our simulations a fast ejecta component as reported in Ref.~\cite{Radice:2018pdn}.

\section{Conclusions}
\label{sec:Conc}
In this work we presented, for the first time, a set of binary neutron star merger simulations that include muons and muonic weak reactions. To do so, we introduced a scheme to produce 4-dimensonal EoS tables parameterized by $(\rho, T, Y_p, Y_\mu)$ by ``dressing'' a 3-dimensional baryonic baseline EoS with a leptonic EoS modeling electrons, positrons, muons and antimuons as relativistic, ideal Fermi gases.

Next, we introduced a scheme for the tabulation of neutrinos opacities and emission rates that, differently from previous works~\cite{Ruffert:1995fs, Ardevol-Pulpillo:2018btx} in which most of the BNS studies are based on, here we included in-medium corrections embodied by the medium-modified $Q$ factor for the charged-current absorption and emission reactions, as in Refs.~\cite{Martinez-Pinedo:2012eaj, Guo:2020tgx, Fischer:2020vie, Ng:2023syk}, but we restricted our treatment to the elastic approximation. In future works it would be important to consider the full kinematics approach of Ref.~\cite{Guo:2020tgx,Fischer:2020vie}, since it implies important corrections to the neutrinos opacities.

We ran a set of BNS simulations adopting the SFHo baseline baryonic EoS, modeling neutrinos as per a leakage scheme incorporating the aforementioned updated sets of neutrinos opacities and emission rates. For comparison purposes, our SFHo$\_$3D setup was simulated with a 3-dimensional EoS including electrons and positrons and the usual 3-species $\{\nu_e, \bar\nu_e, \nu_x\}$. For this particular setup we observed gravitational collapse, in good agreement with Ref.~\cite{Radice:2018pdn}, with smaller ejecta masses by a few tens of percent compared to Refs.~\cite{Radice:2018pdn, Bovard:2017mvn, Lehner:2016lxy}. At this stage, it is difficult to point out to which extent those differences arise due to different hydrodynamics implementations and to neutrinos treatment. 
We also ran three simulations including muons, namely SFHo$\_$4D$\_$3, SFHo$\_$4D$\_$5 and SFHo$\_$4D$\_$5$\_$High, the first one with 3-neutrino species and the remaining with 5-neutrino species $\{\nu_e, \bar\nu_e, \nu_\mu, \bar\nu_\mu, \nu_x\}$, explicitly separating the muon-flavored neutrinos $\nu_\mu,~\bar\nu_\mu$ from the heavy-lepton neutrinos $\nu_x$ and adopting charged-current muonic reactions. The last setup has increased grid resolution with respect to the remaining ones. For SFHo$\_$4D$\_$3 the collapse was delayed by $\sim 7~{\rm ms}$, while for SFHo$\_$4D$\_$5 there is no collapse within our simulation time (up to $20~{\rm ms}$ after the merger) and we note a significant damping of the central rest-mass density oscillations. At increased resolution, SFHo$\_$4D$\_$5$\_$High also does not collapse within $20~{\rm ms}$. We found that the stabilization is provided by increased rotational support. In fact, the softening of the high-density EoS by inclusion of muons leads to a more violent merger event, followed by a stronger centrifugal bounce. This effect is more pronounced for SFHo$\_$4D$\_$5($\_$High), since some muonization in the early post-merger allows the densest regions of the remnant to retain higher $Y_\mu$ compared to SFHo$\_$4D$\_$3, for which $Y_\mu$ is simply advected outwards into the disk. The centrifugal bounce redistributes matter and angular momentum such that the remnant for the muonic runs become less compact and, thus, more stable against gravitational collapse.

Regarding the evolution of the muon content along the post-merger stage we note that when muonic CC reactions are neglected, $Y_\mu$ is advected from the remnant core, distributing $Y_\mu > 0.01$ relatively far within the disk. When muonic CC reactions are included, the disk is found effectively de-muonized, as expected by the $\beta$-equilibration of lower density, cold material and $Y_\mu > 0.01$ is restricted to dense portions of the remnant with $\rho \geq 10^{14}~{\rm g/cm^3}$.

In order to stress out the limitations of our methods, we presented a comparison between opacities obtained with the elastic approximation and the full kinematics approach for selected thermodynamical conditions found in our simulations, concluding that, in general, weak rates in the elastic approximation are underestimated by factors that range between a few and orders of magnitude, depending on the neutrino energy. The difference between approaches is, as expected, more evident at higher densities, where the matter dynamics is less sensitive to details of the neutrino transport scheme. Such a comparison can also be seen in, e.g., Ref.~\cite{Ng:2023syk}.

At small densities ($\rho \leq 10^{10}~{\rm g/cm^3}$) and small temperatures, found in the disk and polar regions, elastic and full kinematics approaches produce almost the same CC rates. Besides, since the opacities there are very small for (anti)muon neutrinos, whose average energies in equilibrium lie in the strongly suppressed region, one does not expect absorption to significantly modify the muonic content. Therefore, we believe that a more realistic neutrino transport treatment should not qualitatively alter the de-muonization scenario presented here, given that it is an expected outcome of $\beta$-equilibration in the freely streaming regime.

We expect, however, that the adoption of a neutrino transport with proper absorption treatment, combined with full kinematics rates, would affect the early post-merger, where temperatures larger than $T = 15~{\rm MeV}$ enable sizable absorption of $\nu_\mu$, which in turn allows those hot spots to retain muons and impact the dynamics of the system by softening the EoS. Additionally, higher-energy (anti)muon neutrinos emitted at higher temperatures could be reabsorbed in optically thinner regions, possibly reducing the de-muonization rates.

For the muonic simulations, we found systematically smaller amounts of ejecta compared to SFHo$\_$3D. This is consistent with the reported higher total neutrino luminosities, which indicate that energy is effectively drawn from outflowing matter elements, preventing them from escaping the system.
Structure-wise, the muonic simulations exhibit less compact and cooler disks. In particular, we noted that the inclusion of muonic CC reactions lead to the suppression of the formation of shocked arms along the disk. Furthermore, despite the initially higher proton fraction in the interior of NSs containing muons, which is retained throughout the post-merger stage, the disk is less protonized than in the non-muonic counterpart.
Overall our results suggest that the inclusion of muons and muonic weak reactions lead to significant consequences regarding the outcomes of BNS merger simulations, mostly affecting the post-merger evolution, the thermal and compositional structure of the remnant and, consequently, the ejecta properties. 
In future works we plan on extending the M1 scheme of Ref.~\cite{Schianchi:2023uky} to account for muon-flavored neutrinos interactions and simulate a number of different baseline baryonic EoSs in order to assess the impacts of a more realistic neutrinos treatment.
Finally, it is important to point out that, at typical BNS remnant temperatures and densities, a sizeable population of pions is expected to be present~\cite{Fore:2019wib}. In this case, the equilibrium composition of a NS is altered by the increase in the proton content, which leads to a softening of the EoS. Consequently, macroscopic properties of a NS are significantly altered by the inclusion of pions~\cite{Vijayan:2023qrt, Pajkos:2025oyf}, as well as BNS merger simulations outcomes, e.g., post-merger GW frequencies, threshold binary mass for prompt gravitational collapse and ejecta properties. Furthermore, weak interaction reactions involving pions and muons may be an important source of opacity for low-energy muon-flavored neutrinos. Hence, we also intend to extend our formalism in the future to investigate to the role pions and muons in BNS merger simulations.

\section{Acknowledgements}
HG, FS and TD acknowledge funding from
the EU Horizon under ERC Starting Grant, no. SMArt-101076369. The simulations were performed on HPE Apollo Hawk at the High Performance Computing (HPC) Center Stuttgart (HLRS) under the grant number GWanalysis/44189, on
the GCS Supercomputer SuperMUC-NG at the Leibniz
Supercomputing Centre (LRZ) [project pn29ba], and on
the HPC systems Lise/Emmy of the North German Supercomputing Alliance (HLRN) [project bbp00049].


%
\end{document}